\documentclass[10pt,twocolumn,letterpaper]{article}

\usepackage[accsupp]{axessibility} 
\usepackage[pagenumbers]{cvpr} 
\usepackage{multirow}
\usepackage{tabularx}

\newcommand{\methodName}{OmniRet\xspace}
\newcommand{\datasetName}{ACM\xspace}

\usepackage{booktabs}
\usepackage{tabularray}
\usepackage{multicol}
\usepackage{multirow}
\usepackage{caption}
\usepackage{graphicx}
\usepackage{listings}
\usepackage{makecell}
\usepackage{enumitem}
\usepackage{comment}
\usepackage{algorithm2e}
\usepackage{algpseudocode}
\RestyleAlgo{ruled}
\UseTblrLibrary{booktabs}

\definecolor{mypink}{rgb}{0.95, 0.95, 1.0}
\definecolor{myorange}{rgb}{0.98, 0.98, 0.98}
\definecolor{gain}{HTML}{34a853}
\definecolor{lost}{HTML}{ea4335}

\newcommand{\Sref}[1]{Sec.~\ref{#1}}
\newcommand{\Eref}[1]{Eq.~(\ref{#1})}
\newcommand{\Fref}[1]{Fig.~\ref{#1}}
\newcommand{\Tref}[1]{Table~\ref{#1}}
\newcommand{\R}[0]{\mathbb{R}}

\newcommand{\x}[0]{\times}
\newcommand{\fst}[1]{\textbf{#1}}
\newcommand{\bst}[1]{{\color{gain}\fst{#1}}}
\newcommand{\snd}[1]{\underline{#1}}
\newcommand{\ndif}[1]{{\color{lost}-#1}}
\newcommand{\pdif}[1]{{\color{gain}+#1}}

\newcommand\mypara[1]{\vspace{1mm}\noindent\textbf{#1}.}

\DeclareMathOperator*{\argmax}{arg\,max}

\definecolor{cvprblue}{rgb}{0.21,0.49,0.74}
\usepackage[pagebackref,breaklinks,colorlinks,allcolors=cvprblue]{hyperref}

\def\paperID{8817} 
\def\confName{CVPR}
\def\confYear{2026}

\title{Efficient and High-Fidelity Omni Modality Retrieval} 

\author{
Chuong Huynh$^{1\dag}$ \quad Manh Luong$^{2\dag}$ \quad Abhinav Shrivastava$^{1}$ \\
\vspace{6pt}{$^1$~University of Maryland, College Park, USA} \hspace{2mm}
{$^2$~Monash University, Australia} \hspace{2mm} {$^\dag$~Equal contribution}\\
Project Page: \url{https://hmchuong.github.io/omniret}
}

\begin{document}
\maketitle
\begin{abstract}
Multimodal retrieval is the task of aggregating information from queries across heterogeneous modalities to retrieve desired targets. State-of-the-art multimodal retrieval models can understand complex queries, yet they are typically limited to two modalities: text and vision. This limitation impedes the development of universal retrieval systems capable of comprehending queries that combine more than two modalities. To advance toward this goal, we present OmniRet, the first retrieval model capable of handling complex, composed queries spanning three key modalities: text, vision, and audio.
Our OmniRet model addresses two critical challenges for universal retrieval: computational efficiency and representation fidelity. First, feeding massive token sequences from modality-specific encoders to Large Language Models (LLMs) is computationally inefficient. We therefore introduce an attention-based resampling mechanism to generate compact, fixed-size representations from these sequences. This shared module is designed to maintain representational diversity and generalization capabilities while remaining sensitive to modality-specific information. Second, compressing rich omni-modal data into a single embedding vector inevitably causes information loss and discards fine-grained details. We propose Attention Sliced Wasserstein Pooling to preserve these fine-grained details, leading to improved omni-modal representations.
OmniRet is trained on an aggregation of approximately 6 million query-target pairs spanning 30 datasets. We benchmark our model on 13 retrieval tasks and a MMEBv2 subset. Our model demonstrates significant improvements on composed query, audio and video retrieval tasks, while achieving on-par performance with state-of-the-art models on others. Furthermore, we curate a new Audio-Centric Multimodal Benchmark (ACM). This new benchmark introduces two critical, previously missing tasks—composed audio retrieval and audio-visual retrieval—to more comprehensively evaluate a model's omni-modal embedding capacity. We believe our benchmark will facilitate the development of universal retrieval systems.
\end{abstract}
\section{Introduction}
\label{sec:intro}

\begin{figure}[t]
    \vspace{-1em}
    \centering
    \begin{subfigure}[t]{\linewidth}
        \centering
        \includegraphics[width=1.0\linewidth]{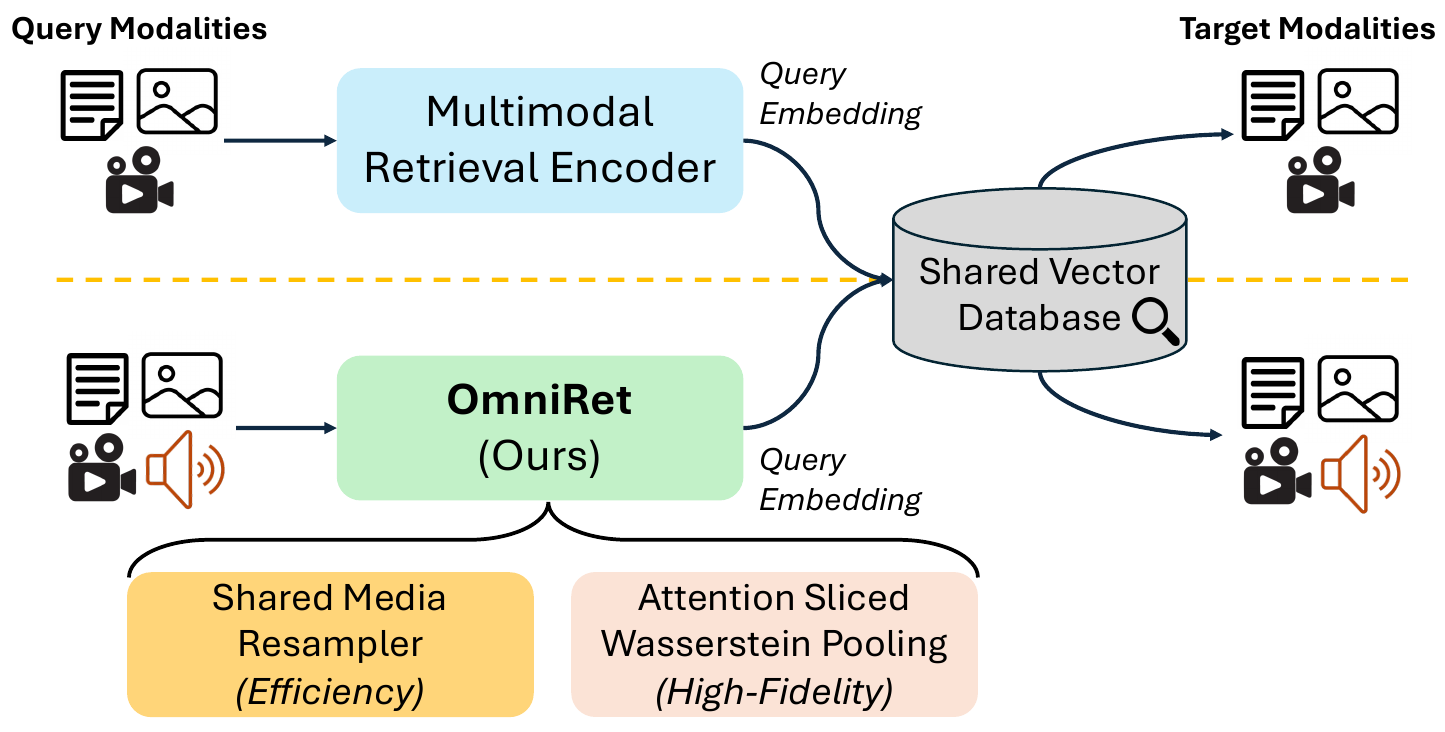}
        \caption{Multimodal Retrieval Systems (top) and our \methodName (bottom)}
        \label{fig:teaser_a}
    \end{subfigure}
    \begin{subfigure}[t]{\linewidth}
        \vspace{0.5em}
        \centering
        \includegraphics[width=1.0\linewidth]{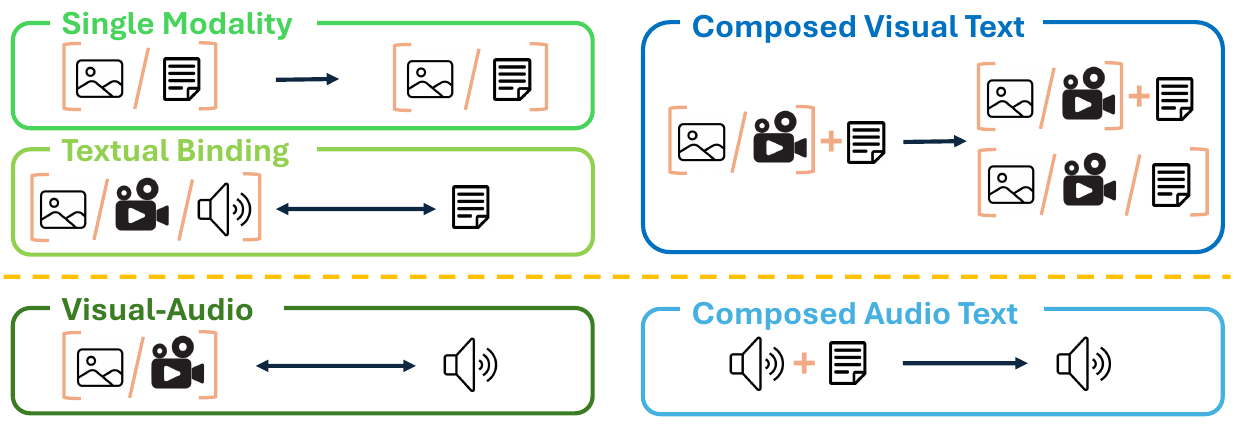}
        \caption{Our evaluation suite with new Audio-Centric MultiModal tasks.}
        \label{fig:teaser_b}
    \end{subfigure}
    \caption{\textbf{(a) Comparison of retrieval systems.} \methodName is the first to handle composed queries from text, vision, and audio. It introduces a Shared Media Resampler for efficiency and Attention Sliced Wasserstein Pooling to preserve high-fidelity, fine-grained details. \textbf{(b) Our  Audio-Centric MultiModal evaluation suite.} We introduce two novel, audio-centric tasks: Visual-Audio Retrieval and Composed Audio-Text Retrieval. These additions fill critical gaps in existing benchmarks, enabling a more comprehensive evaluation of omni-modal retrieval models.}
    \label{fig:teaser}
    \vspace{-1em}
\end{figure}

The landscape of information retrieval has fundamentally shifted from single-modality systems, such as traditional text search, to a complex ecosystem demanding seamless interaction across diverse data types. The proliferation of digital content requires models that can understand heterogeneous, composed queries—for example, finding a target image using a query that combines a related image and a line of text. While pioneering cross-modal models like CLIP~\cite{radford2021learning}, BLIP~\cite{li2022blip,li2023blip}, and CLAP~\cite{wu2023large} have excelled at bi-modal tasks, they fall short in scenarios requiring compositional understanding across multiple modalities. Furthermore, the frontier of omni-modal retrieval, often relying on Multimodal LLMs (MLLMs)~\cite{lin2025mmembed, jiang2025vlmvec,liu2025lamra}, reveals significant gaps. Modalities like video and, most notably, audio remain critically underserved due to a scarcity of dedicated models and comprehensive training data.

The path to a universal retrieval model is obstructed by two primary challenges: \textit{representation fidelity} and \textit{computational efficiency}. On one hand, condensing rich, multi-modal inputs into a single embedding vector creates an information bottleneck that discards the fine-grained details necessary for high-fidelity retrieval~\cite{weller2025theoretical}. Prior works~\cite{lin2025mmembed, jiang2025vlmvec} often resort to simple average pooling or using the `[EOS]' token as the final representation. These methods are efficient, however, the fine-grained information extracted from LLMs could be discarded due to the weighted sum operation. While sophisticated late-interaction mechanisms like ColBERT~\cite{khattab2020colbert} overcome this by retaining token-level embeddings, they introduce prohibitive computational and storage costs, making them impractical for large-scale systems. On the other hand, the sheer scale of modern foundation models presents a significant efficiency challenge. Feeding massive token sequences from media encoders, often exceeding 500 tokens for a single image, into an LLM composer results in an explosion of computational cost. This computational overhead severely limits the feasible batch size, which in turn undermines the effectiveness of contrastive learning, a cornerstone for training state-of-the-art embedding models.

To address these challenges, we introduce \methodName, the first unified framework for omni-modal embedding, capable of handling multimodal queries and targets from three primary modalities: text, vision, and audio (~\Fref{fig:teaser_a}). Our framework makes two primary contributions. First, to tackle the efficiency bottleneck, we introduce an attention-based resampling module that intelligently condenses large sequences of media tokens into a compact, fixed-size set. This module is shared across modalities to enhance generalization, yet maintains modality-specific sensitivity by using separate media latents. We augment this with a novel diversity loss function, ensuring that these condensed tokens retain maximal information and thereby improve the final embedding quality. Second, to achieve high-fidelity representation without sacrificing efficiency, we propose a new pooling method called Attention Sliced Wasserstein Pooling (ASWP), inspired by PSWE~\cite{naderializadeh2021pooling}. By conceptualizing the set of LLM output tokens as a distribution, ASWP computes a rich embedding based on the distance to a set of learnable references~\cite{nguyen2025introduction}. This approach preserves fine-grained, token-level information while maintaining the speed and simplicity of a single-vector system. Our extensive experiments demonstrate that \methodName achieves superior performance on composed retrieval tasks, including those for images, videos, and audio.

We also introduce a new benchmark, the Audio-Centric Multimodal (\datasetName) benchmark (~\Fref{fig:teaser_b}). This benchmark is curated from the VGG-Sound dataset to evaluate universal retrieval systems that extend beyond the traditional text-vision paradigm. We leverage powerful generative models (QwenOmni2.5~\cite{xu2025qwen2} and Gemini2.5~\cite{comanici2025gemini}) to generate high-quality audio captions and modification text describing the relationship between audio pairs. We validate the quality of our synthetic benchmark via human evaluation. Our benchmark contributes two major, previously missing retrieval tasks: composed audio retrieval and audio-visual retrieval.

Our main contribution can be summarized as: \begin{enumerate*}[(i)]
    \item We introduce \methodName, the first universal retrieval framework capable of handling text, vision, and audio. \methodName addresses the critical challenges of computational efficiency and representation fidelity through a novel attention-based resampler and our Attention Sliced Wasserstein Pooling (ASWP) technique.
    \item We propose a large-scale, multi-task training strategy to train \methodName on a massive aggregation of approximately 6 million query-target pairs from 30 distinct datasets.
    \item We curate and will release the \datasetName benchmark, a new audio-centric benchmark featuring two novel tasks, to advance the evaluation of universal retrieval systems.
    \item Through extensive experiments, we demonstrate that \methodName achieves state-of-the-art performance, particularly on composed retrieval tasks across image, video, and audio.
\end{enumerate*}

\section{Related Works}
\label{sec:related_works}
\mypara{Multimodal Embedding} 
Recent advances in multimodal embedding have been driven by large-scale paired datasets and self-supervised learning. Pioneering cross-modal models such as CLIP variants~\cite{radford2021learning, Sun2024EVACLIP18BSC,jia2021scaling} and BLIP~\cite{li2022blip, li2023blip} focus on text-vision alignment, while others like CLAP~\cite{elizalde2023clap, luong2024revisiting} target text-audio modalities. Beyond dataset-centric improvements, several works explore improved pretraining objectives for learning better joint embedding spaces, including SigLIP variants~\cite{zhai2023sigmoid,tschannen2025siglip} and AlignCLIP~\cite{eslami2025mitigate}. More recently, the emergence of LLMs has motivated approaches that leverage their strong representational power~\cite{wang2024unified, gu2025breaking} to construct unified embedding spaces.

\mypara{Universal Multimodal Retrieval} 
Research on universal retrieval systems has rapidly advanced. UniIR~\cite{wei2024uniir} pioneered a universal retriever by training a dual-encoder on 20 datasets, establishing a strong any-to-any benchmark; however, it relies on separate, specialized encoders. Other approaches adapt existing VLMs for embedding tasks~\cite{zhang2024gme,jiang2024e5, jiang2024vlm2vec,meng2025vlm2vec}, achieving strong performance but often limits to text-vision modalities. To improve retrieval, some models~\cite{liu2025lamra,lin2025mmembed} leverage the instruction-following ability of LLMs, using generative rewards~\cite{zhang2025generative, zhou-etal-2025-megapairs} to refine predictions. This line of work is complemented by advances in data curation~\cite{zhang2024magiclens,zhou-etal-2025-megapairs, huynh2025collm} to source high-quality cross-modal pairs for training. Moving beyond two modalities, ImageBind~\cite{girdhar2023imagebind} demonstrated the feasibility of a joint embedding space for six modalities. While this enables omni-retrieval, it highlights a key challenge: managing massive, high-dimensional media inputs from varied encoders. These methods often remain computationally inefficient by directly processing all encoder tokens. Our work directly addresses this efficiency bottleneck.

\mypara{Embedding Pooling}
Embedding pooling techniques are critical for aggregating a sequence of embedding vectors into a single, fixed-size vector~\cite{pham2024composing}. Simple methods, like mean pooling or using the `[EOS]' token's hidden state, are fast but often suboptimal, discarding fine-grained information. More advanced techniques, such as NV-Embed~\cite{lee2025nvembed}, use learnable queries to generate a more descriptive single-vector representation. However, the entire single-vector paradigm faces theoretical limitations~\cite{weller2025theoretical}, spurring the development of late-interaction models. ColBERT~\cite{khattab2020colbert} exemplifies this approach by performing expensive but highly effective token-level comparisons. Subsequent research has focused on making this paradigm more efficient, such as MetaEmbed~\cite{xiao2025metaembed}. While these advancements make late interaction more feasible, they still deviate from the single-vector format required for highly optimized Approximate Nearest Neighbor indexes. To bridge this gap, we introduce a new multimodal-aggregation technique, ASWP, inspired by PSWE~\cite{naderializadeh2021pooling}. Our method captures fine-grained relevance while remaining fully compatible with efficient, large-scale retrieval systems.

\begin{figure*}[t]
    \vspace{-1.5em}
    \centering
    \includegraphics[width=0.8\textwidth]{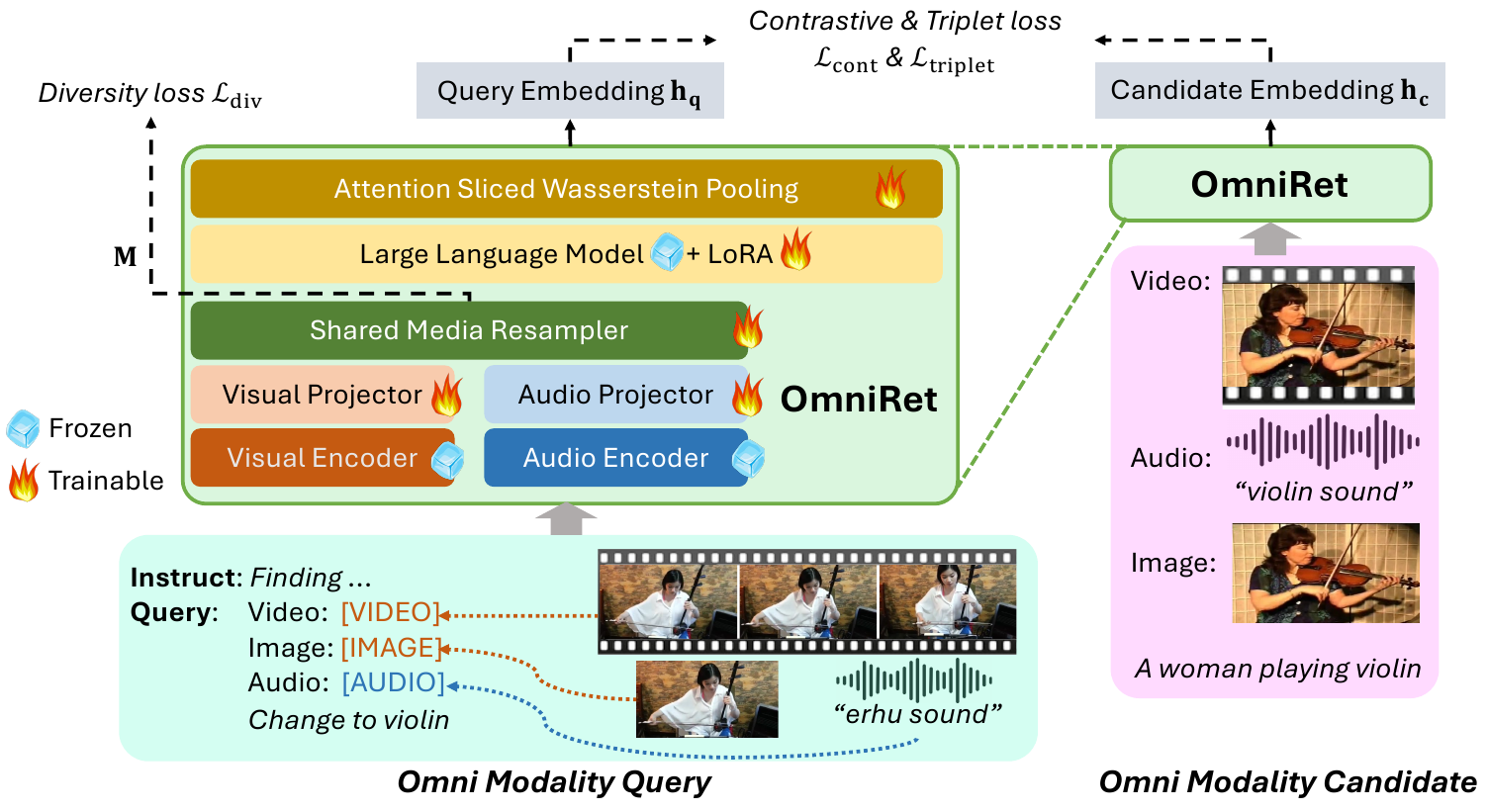}
    \caption{\textbf{Overall \methodName architecture.} Our universal retrieval model integrates specialized visual and audio encoders with a Large Language Model acting as a cross-modal composer. Media inputs are processed through a Shared Media Resampler, and the final embedding is derived via our Attention Sliced Wasserstein Pooling. The model is optimized using Contrastive, Triplet and Diversity losses.}
    \label{fig:omniret:overall}
    \vspace{-1em}
\end{figure*}

\section{\methodName}
\label{sec:method}

\subsection{Problem Definition}

Omni-modal information retrieval aims to match a query $\mathbf{q}$ to a set of candidates $\mathbf{c}$, where both can be of any modality. 
To handle diverse retrieval tasks, we prepend a natural language instruction to the query:
\begin{align}
    \mathbf{q}_\text{inst} = \text{Instruct: } \text{\{\textit{task\_definition}\} } \backslash n \text{ Query: }\{\mathbf{q}\} \label{eq:inst}
\end{align}
where `\{\textit{task\_definition}\}' specifies the retrieval goal and domain. This instruction-tuning approach enhances the model's generalization capabilities. The candidates $\mathbf{c}$, however, remain in their raw, instruction-free form.

Our goal is to train a unified embedding model, $f$, that maps both the instructed query and the candidate to a shared $D$-dimensional space: $\mathbf{h^q}=f(\mathbf{q}_\text{inst})$ and $\mathbf{h^c}=f(\mathbf{c})$. We train $f$ using a Hard-negative InfoNCE loss~\cite{radenovic2023filtering} over in-batch negatives:
\begin{align}
    \mathcal{L}_{\text{cont}} = - \log \frac{e^{\phi(\mathbf{h^q}, \mathbf{h^{c+}})}}{\sum_{c}w(\mathbf{h^q}, \mathbf{h^c})\cdot e^{\phi(\mathbf{h^q}, \mathbf{h^{c}})}} \label{eq:oret:nce}
\end{align}
where $\phi(\mathbf{x},\mathbf{y}) = \frac{1}{\tau} \cos(\mathbf{x}, \mathbf{y})$ is the temperature-scaled ($\tau$) cosine similarity score. Here, $\mathbf{c+}$ and $\mathbf{c-}$ denote positive and negative candidates, respectively. $w(\cdot)$ is an adaptive weighting term for hard-negative mining where $w(\mathbf{h^q}, \mathbf{h^{c+}})=1$ and:
\begin{align}
    w(\mathbf{h^q}, \mathbf{h^{c-}})=\frac{|\mathcal{N}|e^{\beta\phi(\mathbf{h^q}, \mathbf{h^{c-}})}}{\sum_{c-}e^{\beta\phi(\mathbf{h^q}, \mathbf{h^{c-}})}}
\end{align}
with $|\mathcal{N}|$ is the number of negative samples.
Following prior works, we set the temperature $\tau=0.07$ and weighting parameter $\beta=0.5$.

To further providing a strong discriminative learning signal to the model, we also adopt the hinge-based triplet loss
\begin{align}
    \mathcal{L}_\text{triplet} = \sum_{c-} \max [\eta + \phi(\mathbf{h^q}, \mathbf{h^{c-}}) - \phi(\mathbf{h^q}, \mathbf{h^{c+}})]
\end{align}
where $\eta=0.1$ is the margin hyper-parameter.

\subsection{Model Architecture}
Our model, \methodName, employs a Large Language Model (LLM) as a universal composer to process inputs from various modalities, as depicted in \Fref{fig:omniret:overall}. Text inputs are processed directly by the LLM, while other modalities are first encoded by specialized encoders and then projected and resampled into the LLM's token embedding space. All media tokens are then interleaved in the query template defined in \Eref{eq:inst}. Finally, we produce a single embedding vector by aggregating the LLM's output hidden states. The same model is used to encode both the query and the candidate. We keep the media encoders and the LLM frozen, training only our novel modules and a LoRA~\cite{hu2022lora} adapter injected into the LLM. Our novel components are detailed below.

\begin{figure}[t]
    \centering
    \includegraphics[width=\linewidth]{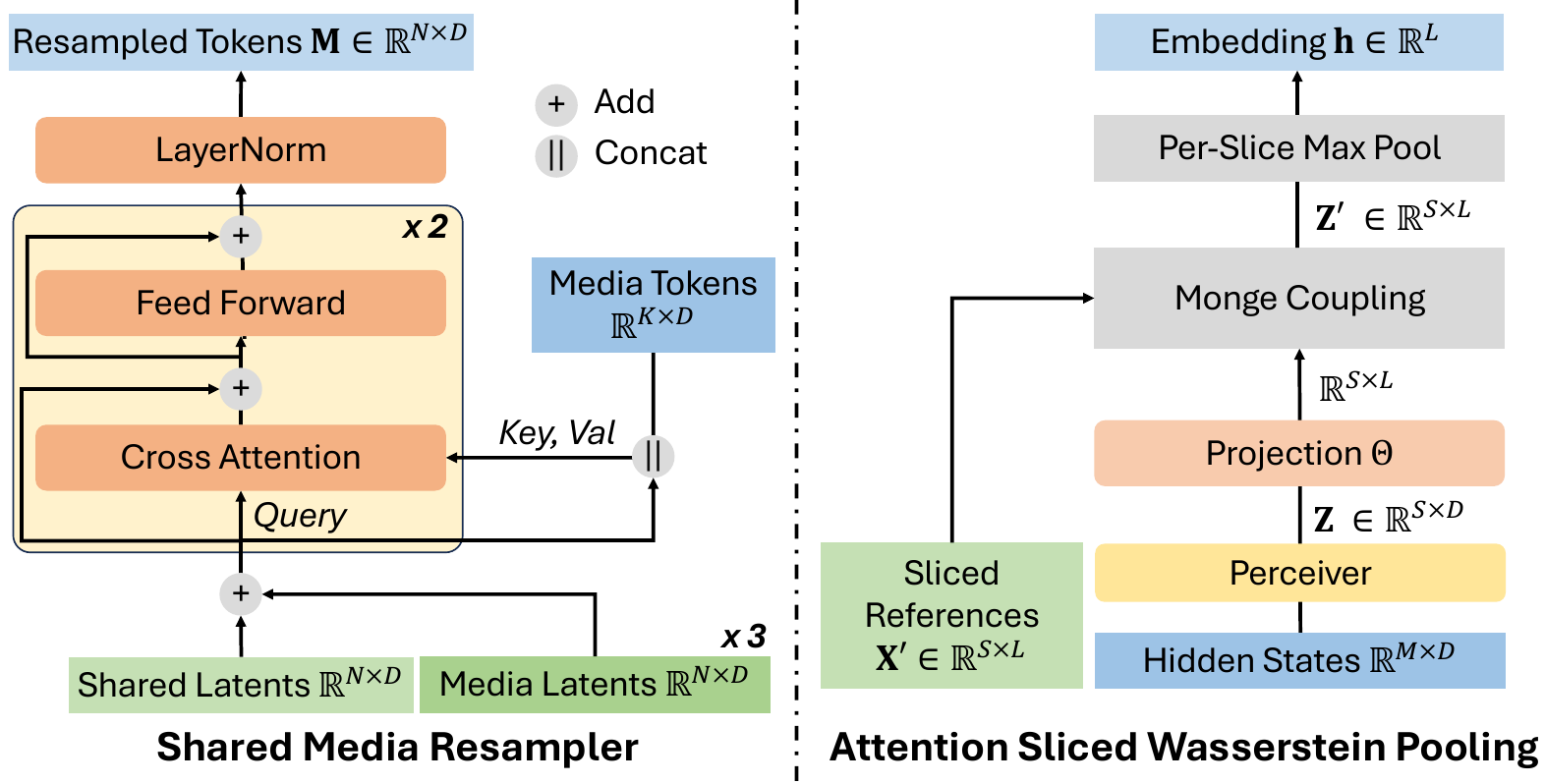}
    \caption{\textit{Left:} Our \textbf{Shared Media Resampler} condenses the output of media encoders into a compact set of latent vectors before they are fed to the LLM. \textit{Right:} Our \textbf{Attention Sliced Wasserstein Pooling (ASWP)} aggregates the final LLM output hidden states into a single, high-fidelity embedding vector. Attention Resampler applies to both output of media encoders and LLM universal encoder.}
    \label{fig:omniret:sampler}
    \vspace{-1em}
\end{figure}

\mypara{Shared Media Resampler}
A key challenge in multi-modal systems is the high token count produced by media (visual and audio) encoders (often $>500$ tokens), which limits training batch sizes and computational efficiency. To address this, we introduce a Shared Media Resampler (\Fref{fig:omniret:sampler}, left). This module is an intermediate layer connecting media tokens with the LLM input space. It condenses a large sequence of media tokens into a small, fixed number of latent vectors using the Perceiver~\cite{jaegle2021perceiver} architecture.

We customize this module to leverage data from different modalities. While a single perceiver module is shared across all media types, we introduce modality-specific latents. These are added to the shared latent queries, enabling the module to adapt to each modality type while retaining its generalization capability. We use all media tokens from the media encoders as input for the resampler, except for videos, where we first apply 3D trilinear interpolation to the video features to reduce frame-level redundancy before resampling. 

\mypara{Diversity on Media Resampled Tokens} 
To ensure the resampled tokens capture diverse information, we apply a diversity regularization loss, $\mathcal{L}_{\text{div}}$, which encourages orthogonality among the output vectors $\mathbf{M} \in \mathbb{R}^{N \times D}$:
\begin{align}
    \mathcal{L}_{\text{div}} = \frac{1}{N^2} \sum \text{smooth}_{L1}\left(\text{Dropout}\left(\max(\mathbf{M}\mathbf{M}^\top, \mathbf{0}) - \mathbf{I}\right)\right)
    \label{eq:loss_div}
\end{align}
Here, we compute the pairwise similarity matrix $\mathbf{M}\mathbf{M}^\top$, clip negative values, and then remove the self-similarity scores (the identity matrix $\mathbf{I}$). Crucially, we apply $\text{Dropout}$ to the resulting matrix before the loss. This acts as a sparse sampling mechanism, ensuring that the loss is computed on only a small, randomly selected subset of token pairs in each training step. This efficiently encourages global diversity.

The $\text{smooth}_{L1}$ function (with $\gamma=0.5$) is then applied. This Huber loss is less sensitive to large similarities (outliers) than a standard $L2$ loss, preventing exploding gradients while still penalizing non-orthogonality. The $\text{smooth}_{L1}$ definition is as follows:
\begin{align}
    \text{smooth}_{L1}(x) = \begin{cases} 
    0.5x^2/\gamma,\quad \text{if } x < \gamma \\
    x - 0.5 * \gamma, \quad \text{otherwise}
    \end{cases}
\end{align}

\mypara{Attention Sliced Wasserstein Pooling} 
To aggregate the LLM output hidden states, we first compress the full sequence into a compact set of $S$ latent embeddings,
$\mathbf{Z}=\{\mathbf{z}_1,\ldots,\mathbf{z}_S \mid \mathbf{z}_i \in \mathbb{R}^D\}$,
using an attention-based resampler identical to the Shared Media Resampler in \Fref{fig:omniret:sampler} (left), applied to the LLM tokens. Instead of average pooling $\mathbf{Z}$, which can blur fine-grained token structure, we use Attention Sliced Wasserstein Pooling (ASWP) to summarize the distribution of $\mathbf{Z}$ relative to a set of learnable reference points
$\mathbf{X}=\{\mathbf{x}_1,\ldots,\mathbf{x}_S \mid \mathbf{x}_i \in \mathbb{R}^D\}$.

Inspired by PSWE~\cite{naderializadeh2021pooling}, ASWP compares $\mathbf{Z}$ and $\mathbf{X}$ across $L$ 1D projections
$\Theta=\{\theta_1,\ldots,\theta_L \mid \theta_i \in \mathbb{S}^{D-1}\}$.
For each projection $\theta_i$, we project both sets onto $\theta_i$ and compute their 1D Monge coupling, producing
\begin{align}
    \mathbf{Z}' = [ \psi_1(\mathbf{X}, \mathbf{Z};\theta_1) ; \ldots ; \psi_L(\mathbf{X},\mathbf{Z};\theta_L) ] \in \mathbb{R}^{S \times L},
\end{align}
where $\psi_i(\cdot) \in \mathbb{R}^S$ measures how the projected token distribution aligns with the projected references. Thus, $\mathbf{Z}'$ can be viewed as a learnable histogram-like descriptor that preserves fine-grained distributional information.

Since $\mathbf{Z}'$ is larger than the target embedding size, ASWP further aggregates it with a hard selection step. For each column $\psi_i$ of $\mathbf{Z}'$, we compute a soft score $\mathbf{y}=\text{softmax}(\psi_i)\in[0,1]^S$ and form a one-hot mask
\begin{align}
    m_i^{\text{hard}} = \text{OneHot}\left(\argmax_{0 \le j < S} \mathbf{y}_j\right).
\end{align}
To backpropagate through this discrete choice, we use the straight-through maximum (STM) estimator:
\begin{align}
    \tilde{m}_i = m_i^{\text{hard}} - \text{StopGrad}(\mathbf{y}) + \mathbf{y}.
\end{align}
In the forward pass, $\tilde{m}_i$ is equivalent to the binary mask $m_i^{\text{hard}}$, while in the backward pass it allows gradients to flow through $\mathbf{y}$. We then apply this selection by Hadamard product, $\mathbf{V}=\mathbf{Z'}\odot \tilde{m}(\mathbf{Z'})$, and obtain the final $L$-dimensional embedding $\mathbf{h}$ by column-wise summation: $\mathbf{h}_i = \sum_{j=1}^S \mathbf{V}_{ji}$.  This procedure is illustrated in \Fref{fig:omniret:sampler} (right). In practice, we learn the sliced references of $\mathbf{X'}$ directly.

\mypara{Final Loss} 
The final loss to train \methodName is the linear combination of $\mathcal{L}_\text{cont},\mathcal{L}_\text{triplet}$ and $\mathcal{L}_\text{div}$:
\begin{align}
    \mathcal{L} = \mathcal{L}_\text{cont} + \mu_1 \mathcal{L}_\text{triplet} + \mu_2 \mathcal{L}_\text{div}
\end{align}
where $\mu_1=1$ and $\mu_2=0.1$ is set in our experiments.

\begin{figure}[t]
    \vspace{-1em}
    \centering
    \includegraphics[width=0.9\linewidth]{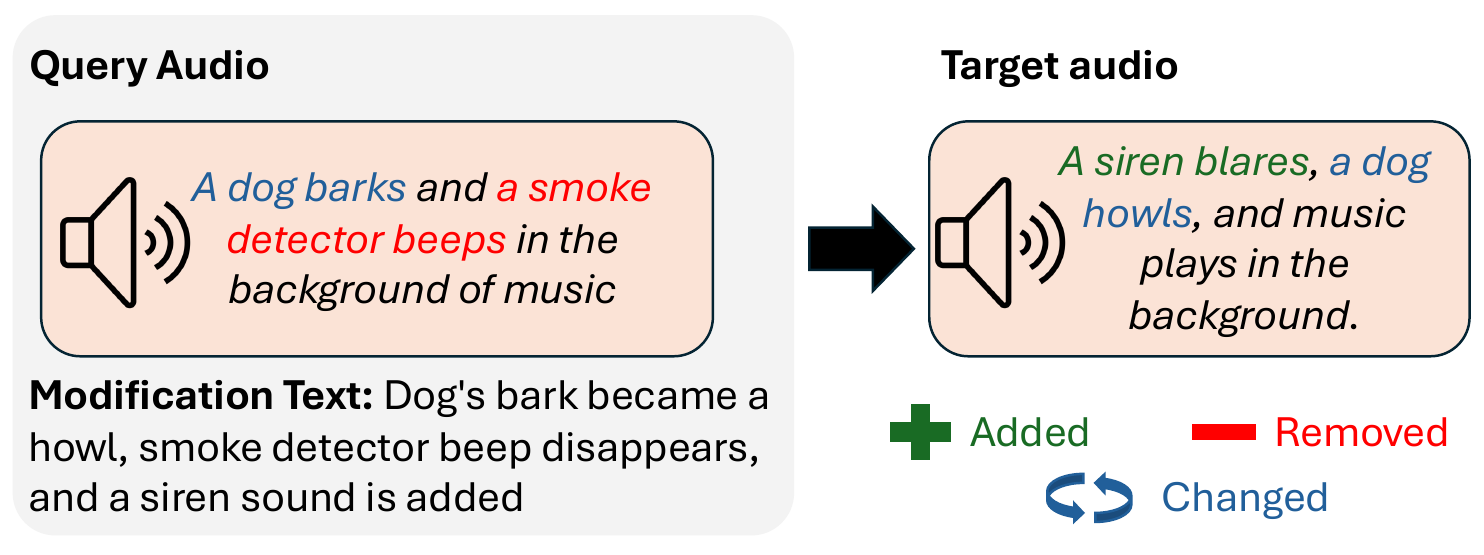}
    \vspace{-10pt}
    \caption{A composed audio retrieval example from our \datasetName benchmark where query composes both audio and text modalities.}
    \label{fig:benchmark_example}
    \vspace{-1em}
\end{figure}
\section{Audio-Centric Multimodal Benchmark}
\label{sec:benchmark}

While composed image retrieval has been extensively investigated~\cite{ray2023cola,wu2021fashion,liu2021image}, no curated benchmark exists for the composed audio retrieval task. Establishing such a benchmark is essential to advance research in complex audio retrieval. Furthermore, few benchmarks exist for audio-to-image or audio-to-video retrieval, largely due to the limited capacity of most models to process such heterogeneous query modalities. To address these gaps, we introduce the Audio-Centric Multimodal (\datasetName) benchmark, which comprises two new tasks: composed audio retrieval (audio-text to audio) and audio-visual retrieval (audio to image/video).


\subsection{Data collection}
\label{sec:data_collection}
\mypara{Audio Source} 
We use the popular VGG-Sound dataset~\cite{chen2020vggsound}, which is widely used for audio-visual tasks. This choice is based on two rationales. First, it contains a large volume of audio-visual correspondence data in the wild, making it well-suited for our audio-visual retrieval task. Second, the composed audio retrieval task requires query-target audio pairs that are sufficiently similar yet exhibit clear distinctions. VGG-Sound, which comprises ~200k 10-second audio clips from $\sim$310 classes, satisfies this requirement.

\mypara{Representative Audio Selection} 
We leverage the ground-truth class labels from VGG-Sound to randomly select 30 audio-video pairs per class. In total, we sample 9,270 pairs to form a representative subset, denoted as $\mathcal{D}$. From this set, we first construct the query-target pairs for the composed audio task, and then reuse these samples for the audio-visual retrieval task.

\begin{table}[t]
\vspace{-1em}
\centering
\small
\caption{\textbf{Subjective evaluation of our \datasetName benchmark.} For the ``Gemini-2.5 w/ text only" method, the audio inputs were replaced with their corresponding generated captions.}
\label{tab:benchmark_analysis}
\footnotesize
\vspace{-0.5em}
\begin{tblr}{width=\linewidth,colspec={@{}X[l]|X[c]@{}},stretch=0}
    \toprule
    Evaluation Method & Accuracy \\ \hline
    Human & 87\% \\ 
    Gemini-2.5 w/ text only & 96\%     \\
    \bottomrule
\end{tblr}
\vspace{-1.5em}
\end{table}

\begin{table*}[t]
    \vspace{-1em}
    \centering
    \caption{\textbf{Performance comparison of \methodName with state-of-the-art baselines across 13 retrieval tasks on the extended version of M-BEIR benchmark.} Our model demonstrates strong, balanced performance on established visual-text tasks while uniquely extending capabilities to the audio modality. Modalities: I (Image), V (Video), A (Audio), T (Text). \fst{Bold} and \bst{green} indicate the best performance in each group and overall, respectively. MMEmbed~\cite{lin2024mm} is included for reference only, as it uses a larger LLM and is not directly comparable.}
    \label{tab:quan_mmbench}
    \scriptsize
    \begin{tblr}{width=\linewidth,colspec={@{}X[2,l]X[1,c]|*{2}{X[1,c]}|*{6}{X[1,c]}|*{5}{X[1,c]}@{}},colsep=2pt,row{13}={mypink},stretch = 0,row{3,9}={myorange},row{11}={fg=gray8}}
    \toprule
    \SetCell[r=2]{l}Model & \SetCell[r=2]{c}LLM Size 
    & \SetCell[c=2]{c}{Uni-Modality} &
    & \SetCell[c=6]{c}{Textual Binding} & & & & &
    & \SetCell[c=5]{c}{Composed Visual Text} & & & & \\
    \hline
    & & I$\rightarrow$I & T$\rightarrow$T &
    I$\rightarrow$T & T$\rightarrow$I & V$\rightarrow$T & T$\rightarrow$V & A$\rightarrow$T & T$\rightarrow$A
    & 
    T$\rightarrow$I,T & I,T$\rightarrow$T & I,T$\rightarrow$I & I,T$\rightarrow$I,T & V,T$\rightarrow$V
    \\
    \hline
    \SetCell[c=15]{c}\textit{Text-Binding Pretrained Models} \\
    CLIP~\cite{radford2021learning} & - & 25.9 & 40.5 
            & 43.1 & 37.5 & 14.2 & 15.9 & - & -
            & - & - & - & - & - \\
    SigLIP~\cite{zhai2023sigmoid} & - & 25.9 & 34.0 
            & 56.5 & 52.1 & 19.7 & 22.5 & - & -
            & - & - & - & - & - \\
    PE-Core~\cite{bolya2025perception} & - & \bst{32.0} & \fst{56.8} 
            & \bst{58.0} & \bst{53.4} & \fst{32.3} & \fst{29.5} & - & -
            & - & - & - & - & - \\
    MMT~\cite{oncescu2021audio} & - & - & - & -
            & - & - & - & 49.8 &  46.8
            & - & - & - & - & - \\
    CLAP~\cite{wu2023large} & - & - & - & -
            & - & - & - & \fst{63.9} & \fst{56.6}
            & - & - & - & - & - \\
    \hline
    \SetCell[c=15]{c}\textit{Multi-Task Finetuning Models} \\
    CLIP$_{SF}$~\cite{wei2024uniir} & - & 28.4 & 83.7
        & 50.0 & 45.1 & - & -  & - & -
        & 63.6 & 41.0 & 26.4 & 60.6 & - \\
    MMEmbed~\cite{lin2024mm} & 7B & 32.1 & 96.7 
        & 49.9 & 44.2 & - & - & - & -
        & 77.2 & 47.2 & 38.5 & 68.5 & - \\
    VLM2VecV2~\cite{meng2025vlm2vec} & 1.5B & \fst{30.0} & 81.1 
        & 43.4 & 39.8 & 17.6 & 18.4  & - & - 
        & 61.6 & 24.5 & 28.7 & 33.6 & 76.4 \\
    OmniRet & 1.5B & 24.4 & \bst{86.7}
        & \fst{50.6} & \fst{46.9} & \bst{43.8} & \bst{43.2} & \bst{66.8} & \bst{62.4} 
        & \bst{70.5} & \bst{44.4} & \bst{36.5} & \bst{64.8} & \bst{86.2} \\
     \bottomrule
    \end{tblr}
    \vspace{-1em}
\end{table*}

\mypara{Query-Target Filtering}
Our filtering process aims to find a set of 3-5 target audios for each query audio $A \in \mathcal{D}$ that are semantically similar but not identical. Since VGG-Sound lacks captions, we first generate a caption for every audio in VGG-Sound using QwenOmni2.5~\cite{xu2025qwen2}.
Let $\kappa$ be the Sentence-BERT~\cite{reimers-2019-sentence-bert} cosine similarity between the caption of $A$ and the caption of another audio in the dataset. To find targets for $A$, we first rank the top 50 audios by $\kappa$ in descending order. From this ranked list, we select up to 5 audios that satisfy three criteria:
\textit{(1) Similarity:} The caption similarity must be in the range $0.6 <\kappa < 0.8$, ensuring both relevance and diversity.
\textit{(2) Textual Diversity:} A candidate is excluded if its $\kappa$ differs by less than 0.05 from the most recently added target, thereby maintaining sufficient distinctiveness among selected samples.
\textit{(3) Acoustic Diversity:} A similar criterion to (2) is applied, but using CLAP~\cite{wu2023large} audio embeddings, ensuring distinctiveness in the acoustic information.
We terminate the selection process for $A$ once 5 targets are found and discard $A$ entirely if fewer than 3 targets can be found. After this filtering process, we retain 1,292 audios from $\mathcal{D}$ as our final query set. Since each of these 1,292 query audios has 3 to 5 associated targets, we obtain a total of 4,251 target audios.

\mypara{Benchmark Construction} 
For each (query, target) audio pair identified, we generate a modification sentence describing the difference between them. We employ the Gemini2.5-Flash-Lite model~\cite{comanici2025gemini} in ``thinking mode", prompting it with the captions of the query-target pair (details in supplementary). This process constructs the composed audio benchmark as a set of triplets: $(\text{query-audio}, \text{modification-text}, \text{target-audio})$. ~\Fref{fig:benchmark_example} shows an example.

Furthermore, we construct the audio-visual retrieval task by reusing the 1,292 audio queries. The benchmark consists of $(\text{audio}, \text{video})$ and $(\text{audio}, \text{image})$ pairs. The target videos are from the original 1,292 query audio-video samples, and the target images are extracted from their middle frames. The remaining audio-video samples from $\mathcal{D}$ (that were not selected as queries) are added to the candidate pool, resulting in 1,292 queries and 5,480 total candidates.

\subsection{Data Quality Analysis}
We conducted a subjective evaluation to verify the quality of our generated captions and modification texts. We tasked three human annotators with answering 100 multiple-choice questions, each created from a randomly sampled query audio. Each question presented the query audio and the modification text; the answer choices consisted of the correct target audio and two random distractor audios from the same candidate subset. As shown in ~\Tref{tab:benchmark_analysis}, human annotators achieved 87\% accuracy, indicating the task is solvable but non-trivial. In contrast, the Gemini-2.5 model (using text-only inputs) achieved 96\% accuracy. These results confirm that our generated audio captions and modification texts are high-quality and sufficiently descriptive for the composed audio retrieval task.



\section{Experiments}
\label{sec:experiment}

\subsection{Implementation Details}

\mypara{Training Datasets} We extend M-BEIR~\cite{wei2024uniir} with additional text, image, video, and audio retrieval datasets. Specifically, we added 7 datasets (MSMarco~\cite{bajaj2016ms}, HotpotQA~\cite{yang2018hotpotqa}, NaturalQuestion~\cite{kwiatkowski2019natural}, PAQ~\cite{lewis2021paq}, StackExchange~\cite{stackexchange}, NLI~\cite{bowman2015large}, SQuAD~\cite{rajpurkar2016squad}) from MTEB~\cite{thakur2021beir} for text retrieval following MMEmbed~\cite{lin2025mmembed}; 3 datasets (LLaVA-558K~\cite{liu2023visual}, CC-CoIR~\cite{ventura2024covr}, MTCIR~\cite{huynh2025collm}) to enhance the interaction between image and text; 5 video and text datasets (TGIF~\cite{li2016tgif}, Charades~\cite{sigurdsson2018charades}, WebVid2M~\cite{bain2021frozen}, PE-Video~\cite{bolya2025perception}, WebCoVR~\cite{ventura2024covr}); 4 datasets for audio and text retrieval (AudioCaps~\cite{kim2019audiocaps}, ClothoV2.1~\cite{drossos2020clotho}, WavText5K~\cite{deshmukh2022audio}, WavCaps~\cite{mei2024wavcaps}) and VGGSound~\cite{chen2020vggsound} for audio-visual retrieval. In total, we use 30 datasets for training, totaling approximately 6.2 million query-candidate pairs. The distribution and details of the training datasets can be found in the supplementary material.

\mypara{Benchmarks and Evaluation Metrics} Besides our proposed \datasetName benchmark, we also evaluate models on an extended version of M-BEIR and a subset of the benchmarks from MMEBv2~\cite{meng2025vlm2vec}. For M-BEIR, we extended it to include additional interactions between the audio-text and video-text modalities. Following MMEBv2, we test each benchmark individually and average the results for each group instead of using a global candidate pool. We use Recall@$k$ with $k=5$ for most benchmarks, except for FashionIQ~\cite{wu2021fashion} and Fashion200K~\cite{han2017automatic}, which use $k=10$. On MMEBv2, we compare performance on classification (CLS), retrieval (RET), and video moment retrieval tasks (MRET). Recall@1 is used as the metric for MMEBv2, while on our proposed benchmark, we report Recall@5. We follow the evaluation protocols of prior works for all models and baselines.

\mypara{Baselines} 
On the extended M-BEIR, we evaluate against dual-encoder models trained on text-binding tasks (\eg, image-to-text) and recent baselines finetuned on multimodal (image and video) data. For MMEBv2 baselines, we gather results for models with LLM backbones smaller than 4B parameters to be comparable with our model. On our benchmark with the audio modality, we choose to compare with ImageBind~\cite{girdhar2023imagebind}, CLAP~\cite{wu2023large}, ViT-Lens~\cite{lei2024vit}, OmniBind~\cite{wangomnibind} and Gemma Embedding~\cite{vera2025embeddinggemma}. While ImageBind, ViT-Lens, and OmniBind are strong baselines for audio-image-video alignment, the Gemma encoding captions from QwenOmni~\cite{xu2025qwen2} serves as an alternative solution for the composed audio retrieval task. Details of each baseline are discussed in the supplementary material.

\begin{table}[t]
    \vspace{-1em}
    \centering
    \caption{\textbf{Generalization performance of \methodName against leading models ($<$7B parameters) on a subset of MMEBv2.} Our model achieves outstanding performance on video tasks. Notably, it maintains competitive performance on image retrieval tasks without being fully fine-tuned on their training sets. \fst{Bold} highlights the best score in each group.}
    \label{tab:quan_mmeb}
    \scriptsize
    \vspace{-0.5em}
    \begin{tblr}{width=\linewidth,colspec={@{}X[2,l]X[1,c]|*{2}{X[1,c]}|*{3}{X[1,c]}},colsep=2pt,stretch = 0,row{13}={mypink},row{3,8}={myorange}}
        \toprule
        \SetCell[r=2]{l}Model & \SetCell[r=2]{c}LLM Size & 
        \SetCell[c=2]{c}Image & &
        \SetCell[c=3]{c} Video  \\
         \hline
         & & CLS & RET
         & CLS & RET & MRET \\
         \hline
         \SetCell[c=7]{c} \textit{Image-Text Embedding Models} \\
         UniME~\cite{gu2025breaking} & 1.5B & 59.0 & 64.9 & - & - & -  \\
         LLaVE~\cite{lan2025llave} & 1.5B & 62.1 & 65.2 & - & - & -\\
         B3++~\cite{thirukovalluru2025breaking} & 1.5B & 67.0 & 70.9 & - & - & -\\
         MetaEmbed~\cite{xiao2025metaembed} & 2B & \fst{68.1} & \fst{71.9} & - & - & -  \\
         \hline
         \SetCell[c=7]{c} \textit{Video-Image-Text Embedding Models} \\
         ColPaliv1.3~\cite{faysse2024colpali} & 2B & 40.3 & 48.1 & 26.7 & 21.6 & 25.5 \\
         GME~\cite{zhang2024gme} & 1.5B & 54.4 & 66.9 & 34.9 & 25.6 & 32.4  \\
         VLM2VecV1~\cite{jiang2025vlmvec} & 1.5B & 58.7 & 65.0 & 33.4 & 20.6 & 33.0 \\
         VLM2VecV2~\cite{meng2025vlm2vec} & 1.5B & \fst{62.9} & \fst{69.5} & 39.3 & 28.8 & 38.5 \\
         OmniRet & 1.5B & 51.7 & 65.3 & \fst{48.6} & \fst{36.5} & \fst{43.3} \\
         \bottomrule
    \end{tblr}
    \vspace{-1.5em}
\end{table}

\mypara{Model Architecture and Training Details} 
At the core of our architecture, we employ GTE-Qwen2-1.5B-Instruct~\cite{li2023towards} as the universal LLM for processing and fusing omnimodal information into a coherent embedding space. Each media input is processed by a different encoder. While visual inputs are handled by SigLIP-SO400M-Patch14-384~\cite{zhai2023sigmoid}, audio files are processed by QwenAudio Encoder~\cite{qwenaudio}. Our entire implementation is built upon the publicly available LLaVA~\cite{liu2023visual} codebase. To balance task performance and maintain a large number of in-batch hard negatives, we customized a data sampler that balances the number of samples between tasks within the same batch and randomly selecting one or two datasets per task. With the majority of parameters remaining frozen, our model has only $\sim$84M trainable parameters in total.

Our training protocol is divided into two distinct stages:
\begin{itemize}
    \item \textbf{Stage 1: Warm-up.} We first warm up the model on a curriculum of simple uni-modality and textual binding retrieval tasks, excluding video-related and composed-query datasets. Only the projectors, resampler, and pooling layer are trained while the LLM is kept frozen. We train the model on 2M samples with the batch size of $2048$, approximately $341$ in-batch samples per task.
    \item \textbf{Stage 2: Fine-tuning.} We continue to train the model on all datasets and tasks for approximately 18M training samples. The video media latents are initialized from the weights of image media latents. Besides trainable modules from stage 1, LoRA with a rank of 16 and alpha of 64 is applied to the LLM. We train with a batch size of $3072$, but select only 4 random tasks per batch and apply gradient accumulation in 2 steps to keep training stable and robust across all tasks.
\end{itemize}
More details on the training hyper-parameters can be found in the supplementary material.

\subsection{Quantitative Results}

\Tref{tab:quan_mmbench} compares our \methodName with other methods, both with and without training on the M-BEIR datasets. The first group includes methods trained only on textual binding tasks; while they show strong performance on those specific tasks, their ability is limited as they cannot handle complex data structures or different modality types. For visual-text retrieval, PE-Core~\cite{bolya2025perception} is presented as the strongest baseline, with leading performance on most tasks. For audio, we compare our model with two strong baselines, CLAP~\cite{wu2023large} and MMT~\cite{oncescu2021audio}. Among models fine-tuned on multi-task retrieval, CLIP$_{SF}$~\cite{wei2024uniir} and MMEmbed~\cite{lin2025mmembed} are fine-tuned on the M-BEIR dataset. VLM2VecV2~\cite{meng2025vlm2vec} is another strong baseline, which is fine-tuned on a broader set of datasets that includes the video modality. In comparison with models of a similar size, \methodName achieves leading performance on all tasks except for image-to-image retrieval, often by a large margin. Notably, on audio and video tasks, \methodName also outperforms all specialized models, even those trained on larger amounts of in-domain data.

We then evaluate \methodName's generalization on a subset of tasks from MMEBv2, shown in \Tref{tab:quan_mmeb}. Since our model was not fine-tuned on the training sets for the Visual Question Answering or Visual Document Retrieval tasks, we only compare performance on classification (CLS) and retrieval (RET) tasks. Our model maintains highly competitive performance on image retrieval, achieving a score (approximately 65) that is within the median range of models fully fine-tuned on this data. For video, \methodName achieves state-of-the-art performance even without being fully trained on all datasets, demonstrating the strong generalization ability of our framework.

\begin{table}[t]
    \vspace{-1em}
    \centering
    \caption{\textbf{Performance comparison on our novel \datasetName benchmark.} \methodName illustrats promising results over baselines on both the composed audio and audio-visual retrieval tasks. Modalities: I (Image), V (Video), A (Audio), T (Text). \fst{Bold} and \snd{underline} indicate the best and second-best performance, respectively.}
    \label{tab:quan_ours}
    \scriptsize
    \vspace{-0.5em}
    \begin{tblr}{width=\linewidth,colspec={@{}X[3.5,l]|*{5}{X[1,c]}},colsep=1pt,stretch = 0,row{7}={mypink}}
        \toprule
        Model & A,T$\rightarrow$A & A$\rightarrow$V & V$\rightarrow$A & A$\rightarrow$I & I$\rightarrow$A\\
        \hline
        QwenOmni~\cite{xu2025qwen2} + Gemma~\cite{vera2025embeddinggemma} & \fst{44.6} & 3.3 & 6.3 & 4.4 & 5.4 \\
        CLAP~\cite{wu2023large} & 16.1 & - & - & - & - \\
        
        ImageBind~\cite{girdhar2023imagebind} & 7.32 & \fst{35.5} & \fst{36.3} & \fst{30.1} & \fst{29.7} \\
        ViT-Lens~\cite{lei2024vit} & 10.1 & - & - & 24.7 & 25.9 \\
        OmniBind~\cite{wangomnibind} & 15.6 & - & - & \snd{30.5} & \snd{29.1} \\
        \methodName & \snd{23.0} & \fst{35.5} & \snd{34.4} & 24.5 & 26.0 \\
        \bottomrule
    \end{tblr}
    \vspace{-1.5em}
\end{table}


Finally, on our new \datasetName benchmark (\Tref{tab:quan_ours}), we compare \methodName with five baselines. For composed audio retrieval (A,T$\rightarrow$A), the strongest baseline is a multi-stage text-to-text retrieval pipeline (using Gemma~\cite{vera2025embeddinggemma} with QwenOmni2.5~\cite{xu2025qwen2}-generated captions). Although it shows promising results, this multi-stage pipeline incurs significant additional computational cost. On this same task, we also show the performance of CLAP~\cite{wu2023large}, ImageBind~\cite{girdhar2023imagebind}, ViT-Lens~\cite{lei2024vit}, and OmniBind~\cite{wangomnibind}with features fused as UniIR~\cite{wei2024uniir}. \methodName outperforms by a large margin while remaining an efficient single-stage method. The Gemma-based pipeline, while strong on the composed audio task, fails to work well on other tasks in our benchmark.
This is likely because converting audio and images/videos to textual descriptions breaks the modality binding between audio and vision, thereby substantially degrading audio-visual retrieval performance.
In contrast, ImageBind, ViT-Lens, and OmniBind have the capacity to align visual and audio embeddings, achieving strong performance on all audio-visual retrieval tasks. However, these models lack the capacity to understand the complex, composed queries in the A,T$\rightarrow$A task. Our model, \methodName, demonstrates a promising ability to align both visual and audio modalities, achieving performance similar to ImageBind on audio-to-video retrieval and remaining competitive on other audio-visual tasks, all while dominating the complex composed retrieval task.

\subsection{Ablation Studies}

\begin{table}[t]
    \centering
    \caption{\textbf{Ablation study of our proposed components.} We report the impact on Average Recall (across 6 tasks, trained on 1M samples) when each component is removed or modified.}
    \label{tab:ablation}
    \scriptsize
    \vspace{-0.5em}
    \begin{tblr}{width=\linewidth,colspec={@{}X[3,l]|X[1,c]X[0.5,c]},colsep=2pt,stretch = 0,row{3,6,11,15,19}={myorange},row{2}={mypink}}
    \toprule
    & Avg. Recall & $\Delta $\\
    \hline
   \textbf{ Our Baseline} & 50.2 & 0.0 \\
    \hline
    \SetCell[c=3]{c} \textit{(a) Embedding Vector} \\
    Single $[$EOS$]$ Vector & 43.4 & \ndif{6.8} \\
    Multi (16) Vectors & 49.8 & \ndif{0.4} \\
    \hline
    \SetCell[c=3]{c} \textit{(b) Number of Projections (L) and References (S) } \\
    $L=1536,S=128$ & 47.5 & \ndif{2.7} \\
    $L=1024,S=128$ & 48.7 & \ndif{1.5} \\
    $L=4096,S=64$ & 48.7 & \ndif{1.5} \\
    $L=4096,S=32$ & 48.1 & \ndif{2.1} \\
    \hline
    \SetCell[c=3]{c} \textit{(c) Pooling in ASWP} \\
    Average Pooling & 20.7 & \ndif{29.5}\\
    Learnable Weighted Sum & 48.5 & \ndif{1.7} \\
    Maximum Pooling & 49.2 & \ndif{1.0} \\
    \hline
    \SetCell[c=3]{c} \textit{(d) Media Resampler} \\
    No Resampler & 46.7 & \ndif{3.5} \\
    Separated Resamplers & 49.9 & \ndif{0.3}\\
    Shared Resampler W/o Media Latents & 49.8 & \ndif{0.4} \\
    \hline
    \SetCell[c=3]{c} \textit{(e) Loss Functions} \\
    W/o $\mathcal{L}_\text{triplet}$ & 49.7 & \ndif{0.5} \\
    W/o $\mathcal{L}_\text{div}$ & 47.1 & \ndif{3.1} \\
    \bottomrule
    \end{tblr}
    \vspace{-1.5em}
\end{table}


We conduct comprehensive ablation studies, training each configuration for ~1M samples and reporting the Average Recall across 6 tasks in \Tref{tab:ablation}. For resource-intensive settings (no resampler and multi-vectors), we halved the batch size and doubled the steps (with gradient accumulation) for a fair comparison.
Our default model uses a Shared Media Resampler, ASWP pooling ($L=4096, S=128$) with the Straight-Through Maximum (STM) estimator, and all three loss functions ($\mathcal{L}_\text{cont}, \mathcal{L}_\text{triplet}, \mathcal{L}_\text{div}$).

\mypara{(a) ASWP Improves Embedding Quality} 
Using a single `[EOS]' vector as the embedding cuts performance by 6.8\%. Late interaction (16 vectors) is better than `[EOS]' but slightly worse than ASWP, likely due to the required smaller batch size limiting hard-negative sampling.

\mypara{(b) More Projectors and References are Beneficial}  
Increasing projections $L$ and references $S$ improves results at the cost of computation. We find $L=4096, S=128$ offers the best tradeoff.

\mypara{(c) STM is the Best for ASWP} 
Average pooling performs poorly, as it cancels out informative positive distances to the references with other negative values. Our STM estimator is the best, outperforming standard max pooling and a learnable weighted sum by enabling better gradient flow.


\mypara{(d) Shared Media Resampler is the Best Choice} 
Removing the resampler degrades performance (due to a smaller batch size). Adding our resampler brings a $+3\%$ recall boost. Using separate resamplers or a shared one without our modality-specific latents both perform worse, showing that sharing data and maintaining specificity is key.

\mypara{(e) Triplet and Diversity Losses are Essential} The result is slightly degraded (0.5\% drop) when removing the $\mathcal{L}_\text{triplet}$ loss. In contrast, performance drops significantly (3.1\% drop) without the $\mathcal{L}_\text{div}$ loss. This highlights the critical importance of maintaining diversity on the resampled media tokens before they are fed to the LLM.

\section{Limitations and Conclusion}
\label{sec:discussion}



Despite promising results, \methodName has clear limitations that open avenues for future work. First, due to resource constraints, we leave scaling the model with a larger LLM backbone and more training data for future investigation, which we expect would significantly boost performance. Second, our work focuses on expanding modalities, whereas other works (\eg, MMEBv2) have expanded retrieval tasks. A key future direction is to train a single, unified model on a large-scale dataset incorporating both more tasks and more diverse input types (\eg, depth maps, 3D point clouds, and speech). Finally, while we introduce a new benchmark for audio retrieval, it could be extended to more complex scenarios, such as retrieval over interleaved, mixed-media documents.

In conclusion, we present a novel approach to omni-modality retrieval where image, video, audio, and text are encoded into a unified embedding space by a single model. Our contributions include:\begin{enumerate*}[(1)]
    \item An innovative architecture that encodes any modality into a shared embedding space, considering both the diversity and efficiency of the extracted information; 
    \item A new Sliced Wasserstein embedding pooling method that overcomes the limitations of coarse-grained, single-vector embeddings;
    \item A new training strategy and dataset combination that improve the model's generalization across many retrieval tasks;
    \item New \datasetName benchmark for evaluating interactions between audio and other modalities.
\end{enumerate*} Notably, \methodName achieves superior results with 12/13 leading performance on extended M-BEIR, outstanding results on video benchmarks on MMEBv2 and promising performance on \datasetName. These advancements collectively push the boundaries of omni-modality retrieval, offering more efficient and effective solutions for real-world applications like RAG or recommendation systems.\vspace{-12pt}
{
    \small
    \bibliographystyle{ieeenat_fullname}
    \bibliography{main,supp}
}

\clearpage
\maketitlesupplementary

In this supplementary material, we provide additional details regarding our proposed \methodName model and the newly introduced \datasetName benchmark, which complement the findings in the main paper.
Specifically, \Sref{sec:add_method} details the instructions used for data generation, the architecture of the Shared Media Resampler, and the pseudo-code for ASWP. \Sref{sec:add_bench} describes the full statistics of our benchmark and the specifics of the baseline settings employed. Finally, \Sref{sec:add_exp} outlines all datasets and baselines utilized in our experiments, including the chosen hyper-parameters for each training stage. We conclude with detailed quantitative results on all benchmarks and provide additional ablation studies on the proposed modules, including a comparative analysis of the computational costs across different models.

\section{Additional Method Details}
\label{sec:add_method}

We leverage the instructions introduced in M-BEIR~\cite{wei2024uniir} and extend them to cover the specific requirements of our new datasets. 
We provide a complete list of all instructions used during model training in the code. Note that the first instruction in this file is the one employed during inference and testing.

We provide further details regarding our architecture. The media projectors within \methodName employ the standard design adopted by LLaVA, utilizing two-layer Multi-Layer Perceptrons with GELU~\cite{hendrycks2016gaussian} non-linear activations interleaved between them. For the shared media resampler, we utilize a module consisting of two cross-attention blocks. The hidden dimension throughout the resampler is kept consistent and equal to the dimension of both the input and output features. In our experiments, we set the number of learnable latent vectors to $N=64$. This constraint was imposed by our available computational resources, as increasing $N$ beyond this limit would require a corresponding decrease in batch size and significantly extend the training time. While a larger $N$ may potentially yield improved performance, we found $N=64$ to be the maximum trainable size within our resource limits. For video processing, we incorporate learnable frame embeddings to capture temporal information. These embeddings are initialized using a standard sinusoidal pattern.

The ASWP layer follows the standard Perceiver architecture for its main components. It consists of two cross-attention blocks where the hidden dimension is consistently maintained to match the input and output dimensionality of the layer. The procedural steps of the ASWP mechanism are formally detailed in Algorithm~\ref{alg:pswe}.

\begin{algorithm}
\caption{Attention Sliced Wasserstein Pooling}
\label{alg:pswe}
\KwData{Pooled LLM Hidden States: $\mathbf{Z}=\{\mathbf{z}_i \in \R^D\}_{i=1}^S$}
\KwResult{Embedding vector: $\mathbf{h} \in \R^L$}
\textbf{Trainable parameters:}\\
\quad Sliced Reference Set: $\mathbf{X}'=\{\mathbf{x}_i \in \R^D\}_{i=1}^S$\;
\quad Slicers (Projectors): $\Theta =\{\theta_l \in \R^D \}_{l=1}^L$\;
\For{$l=1$ to L} {
    Calculate sliced input: $\mathcal{S}^\mathbf{Z}_l=\{\mathbf{z}_i\theta_l\}_{i=1}^S \in \R^{S}$\;
    Sort the distributions: $\pi_\mathbf{Z} = \text{argsort}(\mathcal{S}^\mathbf{Z}_l)$ and $\pi_\mathbf{X} = \text{argsort}(\mathbf{X}')$\;
    Let $\pi_\mathbf{X}^{-1}$: indices that permutes the sorted reference set back to the origin\;
    Compute per-slice embedding: $\psi_l=\{\mathcal{S}^\mathbf{Z}_l[\pi_\mathbf{Z}[\pi_\mathbf{X}^{-1}[i]] - \mathcal{S}_l^\mathbf{X}[i]\}_{i=1}^S \in \R^S$\;
}
Set of embeddings: $\mathbf{Z}'=\{\psi_l\}_{l=1}^L \in \R^{S\x L}$\;
Pooled embedding $\mathbf{h}$ computed via STM;
\end{algorithm}

\section{Additional Benchmark Details}
\label{sec:add_bench}
\begin{table}[h]
    \centering
    \caption{The Statistics of the Audio-Centric Modality Benchmark}
    \label{tab:benchmark_stats}
    \footnotesize
    \begin{tblr}{width=\linewidth, colspec={@{}X[1,l]X[1,c]X[1,c]},stretch=0}
        \toprule
        Task                & No. Queries & No. Candidates \\ \hline
        A,T$\rightarrow$A & 1292     & 4251      \\ \hline
        A$\rightarrow$I & 1292     & 5480      \\ 
        I$\rightarrow$A & 1292     & 5480      \\ 
        A$\rightarrow$V & 1292     & 5480      \\
        V$\rightarrow$A & 1292     & 5480      \\
        \bottomrule
    \end{tblr}
\end{table}

\begin{table*}[ht!]
\centering
\caption{Prompt structure to generate audio captions in the Audio-Centric Multimodal Benchmark.}
\label{tab:prompt_audio_gen}
\begin{tabularx}{0.7\textwidth}{>{\bfseries}l X}
\toprule
System & \texttt{You are a language assistant that helps to generate audio captions.} \\
\midrule
Prompt & \texttt{Help me generate a concise audio caption, which does not exceed 30 words. The generated audio should focus on describing acoustic events within the audio and their temporal relationships. Only generate the caption without explanation.} \\
& \\
& \texttt{[AUDIO]} \\
\bottomrule
\end{tabularx}
\end{table*}

\begin{table*}[ht!]
\centering
\caption{Prompt structure to generate modification texts in the Audio-Centric Multimodal Benchmark.}
\label{tab:prompt_modi_text_gen}
\begin{tabularx}{0.7\textwidth}{>{\bfseries}l X}
\toprule
System & \texttt{You are a language assistant that helps to generate the modification text between two audio captions.} \\
\midrule
Prompt & \texttt{Generate the modification text for the following pair of audio captions:} \\
&\texttt{Caption 1: [CAPTION 1]} \\
& \texttt{Caption 2: [CAPTION 2]} \\
&\\
&\texttt{<Good example>} \\
&... \\
&\texttt{</Good example>} \\
&\\
& \texttt{<Bad example>} \\
&... \\
& \texttt{</Bad example>} \\
& \\
&\texttt{You need to answer the changes from audio 1 to audio 2. The text needs to be concise and details as you can hear the audio, not as you are reading the text. The modification text should not exceed 20 words.
You should not add words "details, specific, description, context, clarify, description, and caption" to the text.
Also, you should not focus on the sound location, only focus on the acoustic sounds and their temporal relationship.
You should focus on the coustic aspect of these captions, which can only be heard, do not focus on the visual aspect which can only be seen, such as "large, small, big, or a specific object/subject name". }\\
\bottomrule
\end{tabularx}
\end{table*}

\begin{figure*}[ht]
    \centering
    \includegraphics[width=0.8\textwidth]{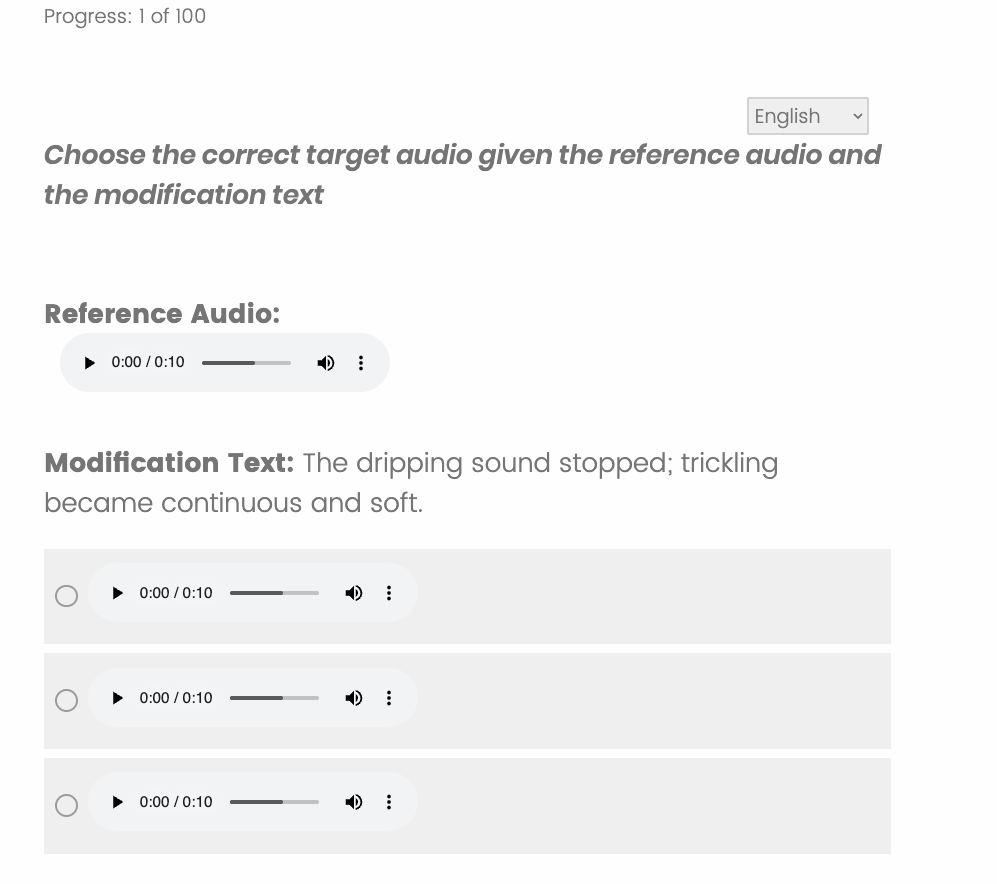}
    \caption{The Qualtric survey for verifying the quality of generated texts in our ACM benchmark.}
    \label{fig:subjetive_exp}
\end{figure*}

The statistics for the five retrieval tasks in the \datasetName benchmark are presented in \Tref{tab:benchmark_stats}. We used the prompts specified in \Tref{tab:prompt_audio_gen} and \Tref{tab:prompt_modi_text_gen} to generate the audio captions and modification texts, respectively. To ensure data quality, a subjective evaluation was conducted using the Qualtrics survey platform to verify the quality of the generated audio captions and modification texts. \Fref{fig:subjetive_exp} illustrates the structure of the survey used for this verification process.

We conduct additional studies to assess the quality of \datasetName on a 300-sample subset. Since the dataset is generated by Gemini 2.5, we use GPT-4o for evaluation to avoid self-preference bias. For each sample, we score naturalness, fluency, and hallucination on a 1--5 scale. \datasetName achieves strong average scores of 4.4, 4.1, and 4.5, respectively. On a clean subset of 250 samples (\Tref{tab:quan_clean}), where all scores are at least 4, OmniRet maintains results consistent with \Tref{tab:quan_ours}, indicating that synthetic artifacts or distribution shifts do not favor our method.

\begin{table}[t]
\centering
\caption{Performance on \datasetName Clean Subset.}
\label{tab:quan_clean}
\scriptsize
\begin{tblr}{width=\linewidth,colspec={@{}X[2,l]|*{5}{X[1,c]}},colsep=2pt,stretch = 0, row{4}={mypink}}
\toprule
Model & A,T $\rightarrow$ A & A $\rightarrow$ V & V$\rightarrow$ A & A$\rightarrow$ I & I$\rightarrow$ A \\ \hline
CLAP~\cite{elizalde2023clap} & 18.4 & -    & -    & - & - \\
ImageBind~\cite{girdhar2023imagebind} & 10.5 & 38.7    & 39.4    & \textbf{33.9}  & \textbf{33.5}  \\
OmniRet     & \textbf{28.2}    & 34.2 & 36.7 & 26.0  & 24.0  \\ \bottomrule
\end{tblr}
\end{table}


To comprehensively evaluate the effectiveness of the \methodName model on the \datasetName benchmark, we compare it against three strong baselines: a text-to-text retrieval method, CLAP, and ImageBind.

\mypara{Text-to-Text Retrieval Baseline} We benchmark against a standard text-to-text retrieval approach using the Gemma embedding model~\cite{vera2025embeddinggemma}. For this baseline, we utilize audio and video captions generated by the Qwen Omni2.5 model~\cite{xu2025qwen2}.

\begin{itemize}
    \item \textit{Composed Audio Retrieval}: The query is formed by concatenating the query audio caption $\mathbf{t}_{\text{cap}}$ with the corresponding modification text $\mathbf{t}_{\text{mod}}$:
    \begin{align}
        \mathbf{q} = \mathbf{t}_{\text{cap}} ||\mathbf{t}_{\text{mod}}
    \end{align}    
    The candidates are the candidate audio captions provided in the benchmark.
    \item \textit{Audio-Visual Retrieval Tasks}: We directly use the audio and video captions generated by the Qwen Omni2.5 model to perform these retrieval tasks.
\end{itemize}

\mypara{CLAP} We also benchmark against the CLAP model~\cite{wu2023large}. We adapt the fusion method from the UniIR framework~\cite{wei2024uniir} to handle the composed audio retrieval task. Specifically, the query is modeled as a linear combination of the query audio embedding and the modification text embedding:
\begin{align}
    \mathbf{h}^\textbf{q}_\text{CLAP} = \lambda \cdot g_A(\mathbf{a}) + (1-\lambda)\cdot g_T(\text{t}_{\text{mod}})\label{eq:fusion}
\end{align}
Here, $g_A(\mathbf{a})$ and $g_T(\text{t}_{\text{mod}})$ represent the embedding vectors from the audio and text encoders of the CLAP model, respectively, with the mixing weight $0<\lambda<1$. We set $\lambda=0.5$ in our experiments to obtain the results reported in Table~\ref{tab:quan_ours}. The candidates are the audio embeddings of the candidate audios, encoded by the CLAP audio encoder. Note that we do not evaluate CLAP on the audio-visual tasks, as the model is not designed to handle visual modalities.

\mypara{ImageBind} Another baseline is the ImageBind model~\cite{girdhar2023imagebind}, which learns a joint embedding space across six different modalities.
\begin{itemize}
    \item \textit{Composed Audio Retrieval}: We follow the identical experimental setup described in~\Eref{eq:fusion} with $\lambda=0.5$.
    \item \textit{Audio-Visual Retrieval Tasks}: We extract embeddings using ImageBind's audio and visual encoders and compute the cross-modal similarities directly within the shared embedding space.
\end{itemize}

\mypara{ViT-Lens and OmniBind} Similar to ImageBind, two recent works, ViT-Lens~\cite{lei2024vit} and OmniBind~\cite{wangomnibind}, learn a joint embedding space across multiple unimodal encoders. We follow the same task setting as ImageBind: for composed audio retrieval, we fuse audio and text embeddings, while for audio-visual retrieval, we directly use their joint embeddings.

\section{Additional Experimental Details}
\label{sec:add_exp}

\subsection{Training Datasets}

\Tref{tab:datasets} provides the comprehensive details of all datasets used for training and testing. In total, the training set comprises $6.4$ million query-candidate pairs, while the testing set involves approximately $300$K queries and a large pool of $5.7$ million candidates.

\begin{table*}[t!]
    \centering
    \caption{Analytics of dataset use to train and test OmniRet.}
    \label{tab:datasets}
    \footnotesize
    \begin{tblr}{width=\linewidth,colspec={@{}X[l]|X[l]|X[c]X[c]|X[c]X[c]@{}},stretch=0}
    \toprule
       \SetCell[r=2]{l}Task & \SetCell[r=2]{l} Dataset & \SetCell[c=2]{c}Training & & \SetCell[c=2]{c} Testing \\
       \hline
       & &  No. Queries & No. Candidates & No. Queries & No. Candidates \\
       \hline
        I$\rightarrow$I & NIGHTS~\cite{fu2023dreamsim} & 15.9K & 15.9K & 2.1K & 40K \\
        \hline
        \SetCell[r=8]{l}T$\rightarrow$T & WebQA~\cite{chang2022webqa} & 16K & 27K & 2.5K & 544K \\
        & MSMarco~\cite{bajaj2016ms} & 100K & 99K & - & - \\
        & HotpotQA~\cite{yang2018hotpotqa} & 84.5K & 84.5K & - & - \\
        & NaturalQuestion~\cite{kwiatkowski2019natural} & 100K & 100K & - & - \\
        & PAQ~\cite{lewis2021paq} & 100K & 100K & - & - \\
        & StackExchange~\cite{stackexchange}  & 100K & 100K & - & - \\
        & NLI~\cite{bowman2015large}  & 100K & 100K & - & - & - \\
        & SQuAD~\cite{rajpurkar2016squad} & 87.3K & 87.3K & - & - \\
        \hline
        \SetCell[r=4]{l}I$\rightarrow$T & VisualNews~\cite{liu2020visual} & 100K & 99.9K & 20K & 537.6K \\ 
        & Fashion200K~\cite{han2017automatic} & 15K & 12.2K & 4.9K & 61.7K \\
        & MSCOCO~\cite{lin2014microsoft} & 113K & 543K & 5K & 24.8K \\
        & LLaVA-558K~\cite{liu2023visual} & 272K & 272K & - \\
        \hline
        \SetCell[r=4]{l}T$\rightarrow$I & VisualNews~\cite{liu2020visual} & 99.9K & 100K & 20K & 542K \\ 
        & Fashion200K~\cite{han2017automatic} & 15K & 48.9K & 1.7K & 202K \\
        & MSCOCO~\cite{lin2014microsoft} & 100K & 72.4K & 24.8K & 5K \\
        & LLaVA-558K~\cite{liu2023visual} & 272K & 272K & -  & - \\
        \hline
        \SetCell[r=4]{l}V$\rightarrow$T & TGIF~\cite{li2016tgif} & 78K & 78K & 33K & 33K \\
        & Charades~\cite{sigurdsson2018charades} & 8K & 8K & 8K & 8K \\
        & WebVid2M~\cite{bain2021frozen} & 500K & 500K & - & - \\
        & PE-Video~\cite{bolya2025perception} & 104K & 45K & - & - \\
        \hline
        \SetCell[r=4]{l}T$\rightarrow$V & TGIF~\cite{li2016tgif} & 78K & 78K & 33K & 33K \\
        & Charades~\cite{sigurdsson2018charades} & 8K & 8K & 8K & 8K \\
        & WebVid2M~\cite{bain2021frozen} & 500K & 500K & - & - \\
        & PE-Video~\cite{bolya2025perception} & 45K & 104K & - & - \\
        \hline
        \SetCell[r=4]{l}A$\rightarrow$T & AudioCaps~\cite{kim2019audiocaps} & 91.3K & 91.3K & 975 & 4.9K \\
        & ClothoV2.1~\cite{drossos2020clotho} & 3.8K & 3.8K & 1K & 5.2K \\
        & WavText5K~\cite{deshmukh2022audio} & 4.3K & 4.3K & - & - \\
        & WavCaps~\cite{mei2024wavcaps} & 403K & 403K & - & - \\
        \hline
        \SetCell[r=4]{l}T$\rightarrow$A & AudioCaps~\cite{kim2019audiocaps} & 91.3K & 91.3K & 4.9K & 975 \\
        & ClothoV2.1~\cite{drossos2020clotho} & 3.8K & 3.8K & 5.2K & 1K \\
        & WavText5K~\cite{deshmukh2022audio} & 4.3K & 4.3K & - & - \\
        & WavCaps~\cite{mei2024wavcaps} & 403K & 403K & - & - \\
        \hline
        \SetCell[r=2]{l}T$\rightarrow$I,T & WebQA~\cite{chang2022webqa} & 17.2K & 17.1K & 2.5K & 403K \\
        & EDIS~\cite{liu2023edis} & 25.9K & 66.2K & 3.2K & 1M \\
        \hline
        \SetCell[r=2]{l}I,T$\rightarrow$T & OVEN~\cite{hu2023open} & 151K & 54K  & 50K & 677K \\
        & InfoSeek~\cite{chen2023can} & 141K & 22.4K & 11.3K & 612K \\
        \hline
        \SetCell[r=4]{l}I,T$\rightarrow$I & FashionIQ~\cite{wu2021fashion} & 16.2K & 16.2K & 6K & 74.4K \\
        & CIRR~\cite{liu2021image} & 26.1K & 15.7K & 4.2K & 21.6K \\
        & CC-CoIR~\cite{ventura2024covr} & 250K & 250K & - & - \\
        & MTCIR~\cite{huynh2025collm} & 208K & 208K & - & - \\
        \hline
        \SetCell[r=2]{l}I,T$\rightarrow$I,T & OVEN~\cite{hu2023open} & 154K & 32.6K & 14.7K & 335K \\
         & InfoSeek~\cite{chen2023can} & 143K & 17.9K & 17.6K & 488K \\
        \hline
        V,T$\rightarrow$V & WebCoVR~\cite{ventura2024covr} & 500K & 500K & 2.6K & 2.6K \\
        \hline
        A$\rightarrow$V & VGGSound~\cite{chen2020vggsound} & 192.5K & 192.5K & - & - \\
        \hline
        V$\rightarrow$A & VGGSound~\cite{chen2020vggsound} & 192.5K & 192.5K & - & - \\
        \hline
        \SetCell[c=2]{c} \textbf{Total} & & 6.4M & 6.4M & 287K &  5.7M \\
        \bottomrule
    \end{tblr}
    
\end{table*}

\subsection{Baselines}

\noindent\textit{CLIP}~\cite{radford2021learning} is a seminal dual-encoder framework trained on a dataset of 400 million image-text pairs using a contrastive language-image pretraining objective. It learns a shared embedding space by maximizing the cosine similarity between matched pairs while minimizing it for mismatched ones, enabling robust zero-shot generalization across various visual tasks.

\noindent\textit{SigLIP}~\cite{tschannen2025siglip} introduces a pairwise sigmoid loss for language-image pretraining, replacing the standard softmax normalization used in contrastive learning. This objective decouples the batch size from the loss computation, allowing for greater scalability and more stable training with larger batch sizes, ultimately yielding superior performance on multimodal retrieval benchmarks.

\noindent\textit{PE-Core}~\cite{bolya2025perception} proposes extracting visual embeddings from intermediate layers of a pretrained vision transformer rather than the final output layer, demonstrating that these deeper, high-quality feature maps provide superior semantic information. The framework is inherently capable of encoding videos by applying its efficient encoder frame-by-frame, ensuring high-fidelity, perceptually rich representations for video understanding and retrieval tasks.

\noindent\textit{MMT}~\cite{oncescu2021audio} is the first work that introduced text-to-audio retrieval tasks on the AudioCaps and Clotho datasets. The MMT framework leverages the dual towers paradigm to perform audio-text retrieval tasks, where the audio encoder is the ResNet18 pretrained on the VGG-Sound dataset and the text encoder is the word2vec model~\cite{mikolov2013efficient}

\noindent\textit{CLAP}~\cite{wu2023large} is a dual-encoder model pretrained with a contrastive learning objective on a large amount of audio-caption pairs, about 630k pairs. Its learned representation demonstrates a strong alignment across audio and text modalities, enabling Zero-shot and supervised audio classification capabilities.

\noindent\textit{CLIP$_{SF}$}~\cite{wei2024uniir} is an instruction-guided multimodal retriever which is further finetuned from the pretrained CLIP model~\cite{radford2021learning} on the training set of the M-BEIR dataset. The model adopts two fusion approaches, score-level and feature-level fusions, to handle eight multimodal retrieval tasks.

\noindent\textit{MM-Embed}~\cite{lin2025mmembed} is the first work to introduce an MLLM-based universal multimodal retrieval framework. It finetunes a strong text retrieval LLM together with a pretrained visual encoder using modality-aware hard negatives and MLLMs as zero-shot rerankers to achieve strong performance for composed retrieval tasks.

\noindent\textit{VLM2Vec-V1}~\cite{jiang2024vlm2vec} is a Vision Language model trained with a contrastive learning objective to generate multimodal representation vectors. It demonstrates a strong embedding space and can handle any composition of image and text to perform composed retrieval tasks on the MMEB benchmark.

\noindent\textit{VLM2Vec-V2}~\cite{jiang2025vlmvec} is the second version of the VLM2Vec-V1 model. It is trained with a flexible and scalable data sampling procedure to achieve a strong generalization across multimodal tasks.

\noindent\textit{UniME}~\cite{gu2025breaking} is the a novel two-stage framework which adopts MLLMs to learn a universal representative embedding space for multimodal downstream tasks. It proposed two key techniques,  textual discriminative knowledge distillation and hard negative enhanced instruction tuning, to achieve robust discriminative and compositional abilities in multimodal retrieval tasks.  

\noindent\textit{LLaVE}~\cite{lan2025llave} is a universal multimodal LLM-based retriever which adapts hardness-weighted contrastive learning to address the high overlap in similarity distribution between positive and negative pairs. This model shows not only a strong performance in the MMEB benchmark but also good generalization in text-video retrieval in a zero-shot manner.

\noindent\textit{B3++}~\cite{thirukovalluru2025breaking} is a novel batch construction pipeline which is able to harvest high-quality batches from a pretrained teacher model for contrastive learning. The retrieval model trained with this pipeline demonstrates superior performance in the MMEB benchmark with a strong teacher model while achieves a good generalization across domains and tasks with weaker teacher models.

\noindent\textit{MetaEmbed}~\cite{xiao2025metaembed} is a universal multimodal framework which employs the Matryoshka MultiVector training technique~\cite{kusupati2022matryoshka} to organize information by granularity across multiple latent vectors. This framework is capable of performing test-time scaling for multimodal retrieval tasks, enabling trade-off between efficiency and effectiveness.

\noindent\textit{ColPaliv1.3}~\cite{faysse2024colpali} is a specialized Vison Language Model trained with late interaction matching objetive for producing high-quality embedding vectors from image of document pages. This model is a simple yet effective approach which outperforms traditional pipelines for document retrieval tasks.

\noindent\textit{GME}~\cite{zhang2024gme} is a general embedding model which is trained on a large amount of synthetic query-target pairs spanning across diverse multimodal tasks. This modal proposes the fused-modal training techniques to enable universal modality retrieval.

\subsection{Training Hyper-parameters and Procedure}

\Tref{tab:hyperparams} summarizes the complete set of hyperparameters used in the two distinct training stages of our \methodName.

\mypara{Stage 1: Warm-up} The first training stage focuses on uni-modality and text-binding tasks, excluding all video-related tasks.
\begin{itemize}
    \item \textit{Data and Batching:} Each batch is constructed to contain samples from 6 tasks spanning 17 unique datasets. This stage utilizes approximately $2$ million total samples.
    \item \textit{Optimization:} We employ a cosine learning rate (LR) scheduler. The LR is warmed up over 100 steps before decreasing to a minimum value of $1 \times 10^{-4}$.
    \item \textit{Hardware:} All experiments were conducted using 64 NVIDIA RTX 2080 Ti GPUs with mixed precision (FP16) enabled for computational efficiency.
\end{itemize}

\mypara{Stage 2: Fine-tuning} Stage 2 continues the training process using the model weights initialized from Stage 1, incorporating all tasks, including the video-related ones.
\begin{itemize}
    \item \textit{Initialization:} Training begins by resuming the model with a constant learning rate of $1 \times 10^{-4}$ for all modules, with no further warm-up applied.
    \item \textit{LLM Fine-tuning (LoRA):} We fine-tune the LLM components using LoRA, setting the rank to $r=16$ and the scaling factor $\alpha=64$.
    \item \textit{Batch Adjustment:} To increase the diversity and quality of training signals, we double the number of intra-dataset samples per batch. This is achieved by decreasing the number of distinct tasks included in each batch while simultaneously increasing the overall batch size.
    \item \textit{Stabilization:} Gradient accumulation is applied throughout this stage to ensure training stability between tasks despite the increased batch size.
    \item \textit{Data Scale:} The model is trained on a substantially larger scale in this stage, using around $18.4$ million total samples.
    \item \textit{Minimum LR:} The minimum learning rate for all modules is set to zero.
    \item \textit{Hardware:} We trained on the same system with 128 GPUs.
\end{itemize}

\begin{table}[t!]
    \centering
    \caption{\textbf{Training Hyper-parameters in Stage 1 and Stage 2}.}
    \label{tab:hyperparams}
    \footnotesize
    \begin{tblr}{width=\linewidth,colspec={@{}X[2,l]|X[1,c]|X[1,c]@{}},stretch=0}
    \toprule
    & Stage 1 & Stage 2 \\
    \hline
    Batch size & 2,048 & 3,072 \\
    No. Tasks & 6 & 15 \\
    No. Datasets & 17 & 28 \\
    No. Dataset/Batch & 6 & 4 \\
    No. Intra-Dataset Samples & $\approx$ 341 & 768 \\
    No. GPUs & 64 & 128 \\
    No. iterations & 1,000 & 6,000 \\
    No. training samples & $\approx$ 2M & $\approx$ 18.4M \\
    Optimizer & \SetCell[c=2]{c} AdamW \\
    Visual Projector LR & 5e-4 & 1e-4 \\
    Audio Projector LR & 5e-4 & 1e-4 \\
    ASWP LR & 5e-4 & 1e-4 \\
    LoRA LR & - & 1e-4 \\
    LoRA Rank & - & 16 \\
    LoRA Alpha & - & 64 \\
    No. Grad. Accum. Steps & 1 & 2 \\
    LR Scheduler & \SetCell[c=2]{c} Cosine \\
    No. Warmup Steps & 100 & 0 \\
    Min LR & 1e-4 & 0 \\
    \bottomrule
    \end{tblr}
    
\end{table}

\subsection{Extended Quantitative Results}

\begin{table}[t]
    \centering
    \caption{\textbf{MMEBv2 Details with VLM2VecV2 and \methodName}. \fst{Bold} indicates the best performance in each dataset while \snd{underline} indices the performance from training on the datasets. }
    \label{tab:quan_mmebv2_extend}
    \footnotesize
    \begin{tblr}{width=\linewidth,colspec={@{}X[1,l]X[1.5,l]X[1,c]X[1,c]@{}},stretch=0}
         \toprule
         Task & Dataset & VLM2VecV2~\cite{meng2025vlm2vec} & \methodName \\
         \hline
            \SetCell[r=10]{l}I-CLS & ImageNet-1K	& \snd{\fst{80.8}} & 57.0\\
            & N24News	& \snd{\fst{72.9}} & 40.6 \\
            & HatefulMemes & \snd{\fst{56.3}} & 49.9 \\
            & VOC2007	& \snd{\fst{85.0}} & 68.1 \\
            & SUN397 & \snd{\fst{71.0}} &	68.7 \\
            & Place365 & 35.9 & \fst{38.3} \\
            & ImageNet-A & \fst{47.4} & 39.1 \\
            & ImageNet-R	& \fst{89.3} & 79.0 \\
            & ObjectNet & \fst{65.2} & 63.6 \\
            & Country211 & \fst{25.2} & 12.8 \\
            \midrule
            \SetCell[r=12]{l}I-RET & VisDial	& \snd{\fst{82.7}} & 49.2 \\
            & CIRR & \snd{\fst{57.5}} & \snd{52.4} \\
            & VisualNews\_t2i & \snd{\fst{74.5}} & \snd{73.5} \\
            & VisualNews\_i2t	& \snd{\fst{78.2}} & \snd{76.5} \\
            & MSCOCO\_t2i & \snd{\fst{75.3}} & \snd{68.2} \\
            & MSCOCO\_i2t	& \snd{\fst{71.4}} & \snd{68.7} \\
            & NIGHTS	& \snd{\fst{68.6}} & \snd{59.3} \\
            & WebQA	& \snd{\fst{90.6}} & \snd{84.9} \\
            & FashionIQ & 19.5 & \snd{\fst{35.9}} \\
            & Wiki-SS-NQ & \fst{66.9} &	41.9 \\
            & OVEN & 64.3 & \snd{\fst{80.1}} \\
            & EDIS & 84.1 & \snd{\fst{92.5}} \\
            \midrule
            \SetCell[r=5]{l}V-CLS & K700 & 38.0 & \fst{45.3} \\
            & SmthSmthV2 & 42.8 &	\fst{52.3} \\
            & HMDB51 & 40.9 & \fst{48.5} \\
            & UCF101 & 60.0 & \fst{ 65.7} \\
            & Breakfast & 14.8 & \fst{31.4} \\
            \midrule
            \SetCell[r=5]{l}V-RET & DiDeMo	& 30.4 & \fst{35.8} \\
            & MSR-VTT	& 28.3 & \fst{38.7} \\
            & MSVD & 48.1 & \fst{64.9} \\
            & VATEX & 26.5 & \fst{33.6} \\
            & YouCook2 & \fst{10.6} & 9.6 \\
            \midrule
            \SetCell[r=3]{l}V-MRET & QVHighlight	& 49.4 & \fst{66.9} \\
            & Charades-STA & 20.2 & \snd{\fst{22.4}} \\
            & MomentSeeker & \fst{40.8} & 40.5 \\
         \bottomrule
    \end{tblr}
\end{table}
The performance breakdown across all datasets is detailed in \Tref{tab:quan_extend}. This table also includes the results achieved by \methodName after the first training stage, demonstrating the impact of the full training curriculum. To evaluate on the video tasks, video media tokens are initialized with the image media tokens within the Shared Media Resampler.

While the model exhibits large performance improvements after the second stage, particularly across new tasks, we observe only a marginal decrease in performance on the original uni-modality tasks as the model adapts to the expanded task complexity. On the text-binding tasks, our method achieves outstanding performance on audio and video modalities, alongside comparable performance to state-of-the-art methods on all image-text tasks. Furthermore, in composed visual-text tasks, \methodName achieves the best score on 6 out of 9 datasets, often establishing a substantial margin over the second-best competitor.

\Tref{tab:quan_mmebv2_extend} presents a focused comparison of our \methodName with the VLM2VecV2 baseline on a subset of the MMEBv2 benchmark. We note that direct comparison on image-text tasks may be compromised due to potential misalignment in the training datasets used by the respective models. Crucially, however, on video-text benchmarks, which primarily employ zero-shot evaluation settings, our model consistently outperforms VLM2VecV2 by a very large margin. This strong performance differential effectively validates the effectiveness of our architecture design and specialized training paradigm for robust video information encoding.

\subsection{Additional Ablation Studies}

We perform three additional ablation studies to rigorously demonstrate the effectiveness of our proposed modules and hyperparameter choices.

We first isolate the contribution of our proposed resampler and pooling design against strong baselines. Extending from ~\Tref{tab:ablation}, we match the batch size of $1024$ (half of the standard size), which is identical to the batch size used for the ``Multi-Vector (Late Attention)" setting. As shown in \Tref{tab:ablation_extend}, our proposed solution, even with this reduced batch size, remains significantly more effective than completely removing the resampler. While the ``Multi-Vector" approach performs slightly better under the same batch size, this experiment validates the crucial role and effectiveness of our pooling and resampler in extracting useful, high-quality information for the retrieval tasks. 

We then investigate the components within our diversity loss, $\mathcal{L}_\text{div}$ (\Eref{eq:loss_div}). We first remove the Dropout mechanism from the loss function calculation. This resulted in a performance decrease of $1.5$ score points. This result suggests that the unregularized diversity loss is overly aggressive when applied uniformly to all token pairs, indicating that some tokens must retain partial similarity to others. Additional experiments were conducted to determine the optimal regression target, comparing $\text{smooth}_{L1}$ against standard $L1$ and $L2$ norms. The use of $\text{smooth}_{L1}$ proved optimal, leading to an increase in recall of over $1.5$ points, confirming its suitability for balancing distance and robustness in the diversity objective. 

To understand the sensitivity of the Shared Media Resampler to feature dimensionality, we analyze the effect of decreasing the number of latents $N$. Instead of the optimal $N=64$ tokens, we tested reduced sizes of $N=32$ and $N=16$. The corresponding performance dropped significantly by $2.1$ and $2.5$ score points, respectively. This clear correlation demonstrates that employing a larger number of media tokens is essential for successfully capturing the fine-grained information required for high-performance multimodal tasks.

\begin{table}[t]
    \centering
    \caption{\textbf{Additional Ablation Studies on Proposed Modules.} We report the impact on Average Recall (across 6 tasks, trained on 1M samples) when each component is removed or modified.}
    \label{tab:ablation_extend}
    \footnotesize
    \vspace{-0.5em}
    \begin{tblr}{width=\linewidth,colspec={@{}X[3,l]|X[1,c]X[0.5,c]},colsep=2pt,stretch = 0,row{2,6}={myorange},row{3,7}={mypink}}
    \toprule
    & Avg. Recall & $\Delta $\\
    \hline
    \SetCell[c=3]{c} Batch size = 1024 \\
    \textbf{Our Baseline} & 49.7 & 0.0 \\
    Multi (16) Vectors & 49.8 & \pdif{0.1} \\
    No Resampler & 46.7 & \ndif{3.0} \\
    \hline
    \SetCell[c=3]{c} Batch size = 2048 \\
    \textbf{Our Baseline} & 50.2 & 0.0 \\
    W/o Dropout in $\mathcal{L}_\text{div}$ & 48.7 & \ndif{1.5} \\
    Replace smooth$_{L1}$ with L1 in $\mathcal{L}_\text{div}$ & 48.0  & \ndif{2.2} \\
    Replace smooth$_{L1}$ with L2 in $\mathcal{L}_\text{div}$& 48.5 & \ndif{1.7} \\
    Sampled media tokens: $N=32$ & 48.1 & \ndif{2.1} \\
    Sampled media tokens: $N=16$ & 47.7 & \ndif{2.5} \\
    \bottomrule
    \end{tblr}
\end{table}

\subsection{Computational Cost}

\Tref{tab:compute} presents the speed and memory usage of various retrieval models evaluated on the MSCOCO Text-to-Image (T$\rightarrow$I) evaluation set. All models were benchmarked on the same NVIDIA RTX A6000 GPU utilizing SDPA (Scaled Dot-Product Attention) for consistent measurement. As expected, dual-encoder architectures, such as CLIP and SigLIP, demonstrate superior inference speed and significantly lower memory utilization. While VLM2Vec-V2 uses less memory than our \methodName, our model achieves a substantial encoding speedup of $3.5\times$ compared to it. The larger memory footprint observed in \methodName is primarily attributed to the Perceiver blocks used to process the media tokens and the LLM hidden states.

\begin{table}[t!]
    \centering
    \caption{\textbf{Speed and Memory Usage of Retrieval Models on MSCOCO T2I.} The performance is measured on RTXA6000 with SDPA attention.}
    \label{tab:compute}
    \footnotesize
    \begin{tblr}{width=\linewidth,colspec={@{}X[1,l]|X[1,c]X[1,c]@{}},stretch=0}
    \toprule
    Model & Speed (samples/s) & Mem (GB) \\
    \hline
    CLIP & 114.35 & 1.62 \\
    SigLIP & 53.43 & 3.36 \\
    PE-Core & 79.05 & 2.54 \\
    CLIP$_{SF}$ & 116.04 & 1.62 \\
    VLM2VecV2 & 10.21 & 4.39 \\
    \methodName & 35.69 & 6.01 \\
    \bottomrule
    \end{tblr}
\end{table}

\begin{table*}[t!]
    \centering
    \caption{\textbf{Details of \Tref{tab:quan_mmbench} with \methodName's Stage 1 included}. Modalities: I (Image), V (Video), A (Audio), T (Text). \fst{Bold} and \bst{green} indicate the best performance in each group and overall, respectively. MMEmbed~\cite{lin2024mm} is included for reference only, as it uses a larger LLM and is not directly comparable.}
    \label{tab:quan_extend}
    \footnotesize
    \begin{tblr}{width=\textwidth,colspec={@{}X[1,l]|X[1.5,l]|*{5}{X[1,c]}|X[1.2,c]X[1.2,c,fg=gray8]*{3}{X[1.2,c]}@{}},stretch=0,colsep=2pt,row{4,7,22}={myorange}}
    \toprule
     \SetCell[r=3]{l}Task & \SetCell[r=3]{l}Dataset & \SetCell[c=5]{c} Text-Binding Pretrained Models &  &  &  &  & 
     \SetCell[c=5]{c} Multi-Task Finetuning Models &  &  &  &  \\ 
     \midrule
    & & CLIP & SigLIP & PE-Core & MMT & CLAP & CLIP$_{SF}$ & MMEmbed & VLM2VecV2 & OmniRet & OmniRet \\ 
    & & \cite{radford2021learning} & \cite{zhai2023sigmoid} & \cite{bolya2025perception} & \cite{oncescu2021audio}& \cite{wu2023large} & \cite{wei2024uniir} & \cite{lin2024mm} & \cite{meng2025vlm2vec} & (stage 1) & (stage 2) \\
    \midrule
    \SetCell[c=12]{c} \textit{Uni-Modality Tasks} \\
    I$\rightarrow$I & NIGHTS & 25.9 & 25.9 & \bst{32.0} & - & - & \fst{28.4} & 32.1 & 30.0 & 27.3 & 25.0 \\ 
    \hline
    T$\rightarrow$T & WebQA & 40.5 & 34.0 & \fst{56.8} & - & - & 83.7 & 96.7 & 81.1 & \bst{91.7} & 87.8 \\ 
    \hline
    \SetCell[c=12]{c} \textit{Text-Binding Tasks} \\
    \SetCell[r=3]{l}I$\rightarrow$T & VisualNews & 42.0 & 43.2 & \bst{50.8} & - & - & \fst{41.4} & 40.4 & 28.7 & 23.5 & 30.3 \\ 
     & Fashion200K & 7.7 & \bst{36.1} & 31.9 & - & - & 18.1 & 18.4 & 12.1 & 28.5 & \fst{29.9} \\ 
     & MSCOCO & 79.7 & 90.2 & \bst{91.3} & - & - & \fst{90.6} & 90.9 & 89.4 & 84.7 & 87.8 \\ 
     \hline
    \SetCell[r=3]{l} T$\rightarrow$I & VisualNews & 44.4 & 42.8 & \bst{48.2} & - & - & \fst{40.7} & 35.4 & 26.9 & 26.3 & 30.2 \\ 
     & Fashion200K & 7.2 & \bst{36.4} & 32.8 & - & - & 15.2 & 16.5 & 13.2 & 28.9 & \fst{30.9} \\ 
     & MSCOCO & 61.0 & 77.0 & \bst{79.4} & - & - & \fst{79.3} & 80.7 & 79.3 & 74.4 & 76.8 \\ 
    \hline
    \SetCell[r=2]{l} V$\rightarrow$T & TGIF & 20.6 & 25.1 & \bst{35.8} & - & - & - & - & 17.3 & 14.7 & \fst{33.8} \\ 
     & Charades & 7.9 & 14.3 & \fst{28.8} & - & - & - & - & 17.8 & 9.9 & \bst{52.0} \\ 
    \hline
    \SetCell[r=2]{l} T$\rightarrow$V & TGIF & 19.8 & 27.5 & \bst{34.2} & - & - & - & - & 18.2 & 14.4 & \fst{33.4} \\ 
     & Charades & 11.9 & 17.6 & 24.7 & - & - & - & - & 18.6 & 10.7 & \fst{50.7} \\ 
    \hline
    \SetCell[r=2]{l} A$\rightarrow$T & AudioCaps & - & - & - & \fst{76.8} & 76.7 & - & - & - & 76.9 & \bst{81.9} \\ 
     & ClothoV2.1 & - & - & - & 22.7 & \fst{51.1} & - & - & - & \bst{52.8} & 52.6 \\ 
    \hline
    \SetCell[r=2]{l} T$\rightarrow$A & AudioCaps & - & - & - & \fst{72.0} & 70.3 & - & - & - & 73.8 & \bst{76.5} \\ 
     & ClothoV2.1 & - & - & - & 21.6 & \fst{42.9} & - & - & - & 50.1 & \bst{50.4} \\ 
     \hline
    \SetCell[c=12]{c} \textit{Composed Visual-Text Tasks} \\
    \SetCell[r=2]{l}T$\rightarrow$I,T & WebQA & - & - & - & - & - & 75.9 & 85.2 & 77.4 & 52.1 & \bst{80.8} \\ 
     & EDIS & - & - & - & - & - & 51.2 & 69.1 & 45.8 & 50.5 & 58.6 \\ 
    \hline
    \SetCell[r=2]{l} I,T$\rightarrow$T & OVEN & - & - & - & - & - & \bst{48.8} & 46.6 & 28.4 & 15.8 & 45.9 \\ 
     & INFOSEEK & - & - & - & - & - & 33.2 & 47.9 & 20.6 & 22.0 & \bst{40.7} \\ 
    \hline
    \SetCell[r=2]{l} I,T$\rightarrow$I & FashionIQ & - & - & - & - & - & 16.5 & 25.5 & 15.2 & 3.4 & \bst{27.3} \\ 
     & CIRR & - & - & - & - & - & 36.3 & 51.5 & 42.1 & 12.1 & \bst{46.8} \\ 
    \hline
    \SetCell[r=2]{l} I,T$\rightarrow$I,T & OVEN & - & - & - & - & - & \bst{69.9} & 71.2 & 44.6 & 32.8 & 69.5 \\ 
     & INFOSEEK & - & - & - & - & - & 51.4 & 65.9 & 22.5 & 25.5 & \bst{58.1} \\ 
    \hline
    V,T$\rightarrow$V & WebCoVR & - & - & - & - & - & - & - & 76.4 & 66.3 & \bst{85.7} \\
    \bottomrule
    \end{tblr}
\end{table*}

\end{document}